\documentclass[a4paper,10pt]{article}
\usepackage{jheppub}

\usepackage[utf8]{inputenc}
\usepackage{amsmath,amssymb,bm,slashed,braket}
\usepackage{dsfont}
\usepackage{amsfonts}
\usepackage{bm,bbm}
\usepackage{graphicx}
\usepackage{slashed} 
\usepackage[dvipsnames,table]{xcolor}
\usepackage[normalem]{ulem}
\usepackage{soul}
\usepackage{siunitx} 
\usepackage{hyperref}
\hypersetup{colorlinks,citecolor= blue,linkcolor= blue, urlcolor=blue}
\usepackage{ulem}
\usepackage{array}
\usepackage{verbatim}
\usepackage{epsfig}
\usepackage{multirow, booktabs}
\usepackage{bbold}
\usepackage[version=4]{mhchem}
\usepackage[capitalise]{cleveref}
\usepackage{makecell}
\usepackage{amssymb}
\usepackage{pifont}
\newcommand{\xmark}{\ding{55}}%

\allowdisplaybreaks[4]
\def\lsim{\mathrel{\raise.3ex\hbox{$<$\kern-.75em\lower1ex\hbox{$\sim$}}}}
\def\gsim{\mathrel{\raise.3ex\hbox{$>$\kern-.75em\lower1ex\hbox{$\sim$}}}}

\newcommand{\tr}{{\rm tr}}
\newcommand{\td}{{\rm d}}

\newcommand{\calL}{ {\cal L} }
\newcommand{\calO}{ {\cal O} }
\newcommand{\calM}{ {\cal M} }
\newcommand{\calS}{ {\cal S} }
\newcommand{\pve}{{\pmb{v}_{\rm el}^\perp}}
\newcommand{\pSx}{\bm{S}_x}
\newcommand{\pSxt}{ \pmb{\tilde{\cal S}}_x}
\newcommand{\pSe}{\bm{S}_e}
\newcommand{\pr}{{\bm{r}}}

\title{A systematic investigation on dark matter-electron scattering in effective field theories}

\author[a,b]{Jin-Han Liang}
\emailAdd{jinhanliang@m.scnu.edu.cn}
\affiliation[a]{Key Laboratory of Atomic and Subatomic Structure and Quantum Control (MOE), 
Guangdong Basic Research Center of Excellence for Structure and Fundamental Interactions of Matter, 
Institute of Quantum Matter, South China Normal University, Guangzhou 510006, China}
\affiliation[b]{Guangdong-Hong Kong Joint Laboratory of Quantum Matter, 
Guangdong Provincial Key Laboratory of Nuclear Science, 
Southern Nuclear Science Computing Center, South China Normal University, Guangzhou 510006, China}
\author[a,b]{Yi Liao}
\emailAdd{liaoy@m.scnu.edu.cn}
\author[a,b]{Xiao-Dong Ma}
\emailAdd{maxid@scnu.edu.cn}
\author[a,b]{Hao-Lin Wang}
\emailAdd{whaolin@m.scnu.edu.cn}

\abstract{
In this paper, we systematically investigate the general dark matter-electron interactions within the framework of effective field theories (EFTs). We consider both the non-relativistic (NR) EFT and the relativistic EFT descriptions of the interactions with the spin of dark matter (DM) up to one, i.e., the scalar ($\phi$), fermion ($\chi$), and vector $(X)$ DM scenarios. We first collect the leading-order NR EFT operators describing the DM-electron interactions, and construct especially the NR operators for the vector DM case. Next, we consider all possible leading-order relativistic EFT operators including those with a photon field and perform the NR reduction to match them onto the NR EFT. Then we rederive the DM-bound-electron scattering rate within the NR EFT framework and find that the matrix element squared, which is the key input that encodes the DM and atomic information, can be compactly decomposed into three terms. Each term is a product of a DM response function $(a_{0,1,2})$, which is essentially a factor of Wilson coefficients squared,  and its corresponding generalized atomic response function ($\widetilde W_{0,1,2}$). Lastly, we employ the electron recoil data from the DM direct detection experiments (including XENON10, XENON1T, and PandaX-4T) to constrain all the non-relativistic and relativistic operators in all three DM scenarios. We set strong bounds on the DM-electron interactions in the sub-GeV region. Particularly,
we find that the latest PandaX-4T S2-only data provide stringent constraints on dark matter with a mass greater than approximately 20 MeV, surpassing those from the previous XENON10 and XENON1T experiments.

}


\begin{document} 

\maketitle
\setcounter{page}{2}

\section{Introduction}

Although there are substantial pieces of evidence for the presence of dark matter (DM) from astrophysical observations, the particle nature of DM remains unknown. One of the theoretically favored candidates is the weakly interacting massive particle (WIMP), which possesses the necessary properties to account for the dark matter puzzle and also has a potential for detection \cite{Arcadi:2017kky,Roszkowski:2017nbc,Young:2016ala}. 
During the past two decades, significant efforts have been dedicated to searching for WIMPs in DM direct detection experiments \cite{PandaX-II:2017hlx,XENON:2023cxc,LZ:2022lsv,DarkSide:2018bpj}. Despite these efforts, no positive signals have been found, leading to stringent constraints on the DM-nucleon cross section in the weak scale WIMP scenarios \cite{Roszkowski:2017nbc,Bottaro:2021snn, Schumann:2019eaa}. 
An exception to this is light DM below ${\cal O}(10\,\rm GeV)$ mass region, which is still less constrained. Conventional nuclear recoil searches are not sensitive to such low mass DM, making electron recoils an alternative strategy due to the significantly smaller mass of the electron compared to a typical nucleus \cite{LZ:2023poo,XENON:2019gfn,XENON:2021qze,XENON:2022ltv,PandaX:2022xqx,DarkSide:2018ppu,PandaX-II:2021nsg}. 

Since the incoming DM and target nucleons/electrons are non-relativistic, the DM-nucleus and DM-atom scattering can both be well described within the non-relativistic (NR) quantum mechanics framework. To formulate the scattering rate without involving too much DM model detail, a convenient way is to work with an NR effective field theory (EFT). The theory preserves rotational and Galilean invariance, and the most general NR EFT operators can be constructed under the guidance of these symmetries. Concerning the DM direct detection experiments, the NR operators for scalar and fermion DM-nucleon/electron scattering have been systematically investigated in \cite{Fan:2010gt,DelNobile:2018dfg,Fitzpatrick:2012ix}. For the vector DM case, the NR interactions induced in some simplified models were discussed in \cite{Catena:2019hzw,Dent:2015zpa}. Ref.\,\cite{Gondolo:2020wge} studied the NR operators for DM with arbitrary spin within the one-nucleon current approximation. The NR interactions are a convenient and effective description at around the experimental scale; to learn more about their relativistic origin, one can consider the relativistic DM EFT or specific UV models and establish their connection. 

In the relativistic framework the low energy effective field theory (LEFT) extended by a light DM particle with a mass smaller than the electroweak scale $\Lambda_{\tt EW}$ has been widely used in the DM direct detection community. Within this LEFT-like framework the interactions between the DM and SM particles satisfy the SM unbroken symmetries, $\rm SU(3)_{\rm c}\times U(1)_{\rm em}$.
Effective operators for scalar and fermion DM interactions with the quarks, gluons, and photon have been constructed up to dimension 7 (dim 7) \cite{Brod:2017bsw}. The constraints on these EFT operators from DM direct detection experiments, including nuclear recoil and Migdal effects, have been systematically studied \cite{Kang:2018rad,Tomar:2022ofh,Liang:2024tef}.
Furthermore, a global fit for EFT operators of fermion DM has been performed by simultaneously taking into account the constraints from the DM relic abundance, direct and indirect detection, as well as LHC searches \cite{GAMBIT:2021rlp,Beniwal:2022rde}. The nonperturbative contributions for fermion DM-quark tensor interactions are discussed in \cite{Liang:2024tef}. For higher spin DM scenarios, Refs.\,\cite{Catena:2019hzw,Dent:2015zpa} studied nuclear recoil signals in some simplified models involving vector DM, and a systematic study within the LEFT-like framework for vector DM is left for the upcoming work \cite{liang:2024nr}. Moving to the electron recoil case,  
Ref.\,\cite{Bertuzzo:2017lwt} studied the (axial-) vector  four-fermion DM-electron couplings in terms of electron recoils. Refs.\,\cite{Dedes:2017shn,Foldenauer:2018zrz,Bernal:2019uqr,Fiaschi:2019evv} investigated the electron recoil in specific ultraviolet (UV) models, while Refs.\,\cite{Knapen:2017xzo,Darme:2017glc,Baxter:2019pnz,Bloch:2020uzh,XENON:2021qze,Catena:2022fnk} worked with simplified models. Given the well-studied nuclear recoil processes in DM direct detection, it is necessary to conduct a more systematic investigation of DM-electron scattering within the LEFT-like framework. Furthermore, if the full standard model symmetries $\rm SU(3)_{\rm c}\times SU(2)_{\rm L}\times U(1)_{\rm Y}$ are adopted, the LEFT-like framework can be extended to its counterpart of the standard model effective field theory (SMEFT). The higher dimensional operators describing the interactions between DM and the SM heavy particles ($W,Z,h,t$) have also been constructed for different DM scenarios in recent years \cite{Liao:2016qyd,Criado:2021trs,Aebischer:2022wnl,Song:2023jqm}.

The rate of DM-electron scattering in the atomic target is theoretically described by the ionization form factor or atomic response functions and the associated DM response functions. The DM properties are encoded in the DM response functions, which are actually the factors of Wilson coefficients squared and dependent on the DM models considered. The ionization form factor or atomic response functions parameterize the effects of bound electrons, which cause the calculation to be different from that for scattering off a free electron. 
The ionization form factor for the liquid xenon target was first calculated in \cite{Dedes:2009bk,Kopp:2009et,Essig:2011nj,Essig:2012yx} based on some simplified DM models.\footnote{It is also named atomic $K$-factor in some literature \cite{Roberts:2016xfw,Ge:2021snv}.}
Subsequently, the atomic response functions in the liquid xenon and argon targets were examined in \cite{Catena:2019gfa} based on a more general framework of NR EFT, in which three new atomic response functions were introduced in addition to the usual ionization form factor.
The relativistic correction and many-body effect of the ionization form factor were discussed in \cite{Roberts:2015lga,Roberts:2016xfw,Pandey:2018esq}, which could be sizable for a large transfer momentum. In a recent work \cite{Liang:2024lkk} we pointed out a crucial minus sign difference from \cite{Catena:2019gfa} for the vector form factor $\bm{f}_{1\to2}$ when transforming from momentum space to configuration space. This sign correction will affect the phenomenological implications of interactions with electron or DM axial-vector or tensor currents. 

In this work, we systematically investigate DM-electron scattering in atomic targets within the framework of EFTs. We begin by providing the most general NR operators basis for DM-electron interactions up to spin-one DM, which are potentially induced by Lorentz invariant interactions in various DM scenarios. Then we go upwards to the LEFT-like framework, in which the leading-order (LO) operators for the DM-electron and DM-photon interactions are considered. We perform the NR reduction at the leading nonvanishing order to match the Lorentz invariant operators to linear combinations of the NR operators. This will help to constrain the relativistic EFT operators from the NR formalism entering the scattering rate. To describe the DM-atom scattering as generally as possible, we then formulate the scattering rate using the full 24 LO NR EFT operators to fill the details for the claims given in \cite{Liang:2024lkk}. The key input for the scattering rate is the spin-averaged amplitude squared. We will find it can be organized into three terms,  each being a product of a DM response function $(a_{0,1,2})$ and its corresponding generalized atomic response function ($\widetilde W_{0,1,2}$). 
Finally, based on the theoretical framework, we study the numerical sensitivity on the EFT operators by employing the electron recoil data from the XENON10, XENON1T, and PandaX-4T. We constrain both the nonrelativistic and relativistic operators for all three DM scenarios with DM mass ranging from MeV to GeV scale. The constraints on the Wilson coefficients of the EFT operators consistently follow the power counting and their behavior can be well understood by the new formalism we develop. Among the experiments, we find that the PandaX-4T sets the most stringent constraints for DM mass greater than approximately 20 MeV, while XENON10 is more sensitive to DM mass smaller than 20 MeV.

The paper is organized as follows. We first provide the most general NR operators for DM-electron interactions up to spin one in \cref{sec:NRop}. In \cref{sec:NRreduct}, we collect the LO operators for the DM-electron and DM-photon interactions in the LEFT-like framework and match them onto the NR EFT operators in the non-relativistic limit. \cref{sec:scatteringrate} is dedicated to the theoretical formalism for the DM-atom scattering from complete NR EFT interactions. In \cref{sec:expcon}, we explore the constraints on both NR EFT and relativistic EFT interactions from current DM direct detection experiments, including XENON10, XENON1T, and PandaX-4T. Our summary is given in \cref{sec:conclusion}. The calculation details are collected in Appendices:  
\cref{app:NRop} gives a detailed construction of NR operators involving the rank-two tensor $\pmb{\tilde\calS}_x$ in the vector DM case, \cref{app:dmRF} illustrates the calculation of the matrix element squared from the complete NR EFT operators, and the detailed computation of the atomic form factors is provided in \cref{app:electronRF}.

\section{NR operators for DM-electron interactions up to spin one}
\label{sec:NRop}

\begin{table}[!h]
\center
\resizebox{\linewidth}{!}{
\renewcommand\arraystretch{1.3}
\begin{tabular}{| l | c | c | c| c| c| }
\hline
\multicolumn{1}{|c|}{ \multirow{2}*{NR operators} } & \multirow{2}*{Refs.~\cite{Catena:2019gfa,Catena:2022fnk}} 
& \multirow{2}*{Power counting} & \multicolumn{3}{|c|}{DM type} 
\\ 
\cline{4-6}
& & & scalar & fermion & vector
\\
\hline
$\calO_1=\mathds{1}_x \mathds{1}_e$ & \checkmark & 1  &  \checkmark &  \checkmark &  \checkmark 
 \\
\hline
$\calO_3= \mathds{1}_x \left({i \bm{q} \over m_e}\times \pve \right) \cdot  \pSe$ 
&  \checkmark & $\color{cyan} q v$ &  \checkmark &  \checkmark &  \checkmark 
\\
\hline
$\calO_4 = \pSx \cdot \pSe$ &  \checkmark & 1 &  $-$ &  \checkmark &  \checkmark 
\\
\hline
$\calO_5 = \pSx \cdot \left({i\bm{q} \over m_e}\times \pve\right) \mathds{1}_e$ 
&  \checkmark & $\color{cyan} q v$ &  $-$ &  \checkmark &  \checkmark 
\\
\hline
$\calO_6 =  \left(\pSx\cdot {\bm{q}\over m_e}\right) \left({\bm{q}\over m_e} \cdot \pSe \right)$ 
&  \checkmark & $\color{cyan} q^2$ &  $-$ &  \checkmark &  \checkmark 
\\
\hline
$\calO_7 = \mathds{1}_x \, \pve \cdot \pSe$ &  \checkmark & $\color{magenta} v$ &  \checkmark &  \checkmark &  \checkmark 
\\
\hline
$\calO_8 = \pSx \cdot \pve \, \mathds{1}_e$ &  \checkmark & $\color{magenta} v$ &  $-$ &  \checkmark &  \checkmark 
\\
\hline
$\calO_9 = -\pSx \cdot\left({i\bm{q}\over m_e}\times \pSe \right)$ &  \checkmark & $\color{magenta} q$ &  $-$ &  \checkmark &  \checkmark 
\\
\hline
$\calO_{10} =  \mathds{1}_x \, {i \bm{q}\over m_e}\cdot \pSe $ & \checkmark & $\color{magenta} q$ &  \checkmark &  \checkmark &  \checkmark 
\\
\hline
$\calO_{11} =  \pSx \cdot {i\bm{q}\over m_e} \mathds{1}_e$ & \checkmark & $\color{magenta} q$ &  $-$ &  \checkmark &  \checkmark 
\\
\hline
$\calO_{12} = - \pSx \cdot (\pve\times \pSe)$ &  \checkmark & $\color{magenta} v$ &  $-$ &  \checkmark &  \checkmark 
\\
\hline
$\calO_{13} = (\pSx \cdot \pve) \left({i\bm{q}\over m_e}\cdot \pSe  \right)$ &  \checkmark & $\color{cyan} qv$ &  $-$ &  \checkmark &  \checkmark 
\\
\hline
$\calO_{14} = (\pSx \cdot {i\bm{q} \over m_e} ) (\pve \cdot \pSe)$ & \checkmark & $\color{cyan} qv$ &  $-$ &  \checkmark &  \checkmark 
\\
\hline
$\calO_{15} =\pSx \cdot {\bm{q}\over m_e}\left[{\bm{q}\over m_e}\cdot(\pve \times \pSe)\right] $  
 & \checkmark & $\color{orange} q^2 v$ &  $-$ &  \checkmark &  \checkmark 
\\
\hline
\cellcolor{gray!25}$\calO_{17} = {i \bm{q}\over m_e} \cdot \pSxt \cdot \pve \,\mathds{1}_e$ 
& ${1\over 3} {i\bm{q}\cdot\pve \over m_e} \calO_1 - \calO_{17}'$  & $\color{cyan} qv $ &  $-$ &  $-$ &  \checkmark 
\\
\hline
\cellcolor{gray!25}$\calO_{18} = {i \bm{q} \over m_e} \cdot \pSxt \cdot  \pSe$ 
&  $ {1\over 3} \calO_{10} - \calO_{18}'$ & $\color{magenta} q $ &  $-$ &  $-$ &  \checkmark 
\\
\hline
\cellcolor{gray!25}$\calO_{19} = {\bm{q}\over m_e} \cdot \pSxt \cdot{\bm{q}\over m_e} \mathds{1}_e$ 
& ${1\over 3} {\bm{q}^2\over m_e^2} \calO_1 - \calO_{19}'$ & $\color{cyan} q^2 $ &  $-$ &  $-$ &  \checkmark 
\\
\hline
\cellcolor{gray!25}$\calO_{20} = - {\bm{q} \over m_e} \cdot \pSxt \cdot \left({\bm{q}\over m_e} \times  \pSe\right)$ 
& $-\calO_{20}'$ & $\color{cyan} q^2$ &  $-$ &  $-$ &  \checkmark 
\\
\hline
\cellcolor{gray!25}$\calO_{21} = \pve\cdot\pSxt\cdot\pSe $ &  \xmark & $\color{magenta} v$  &  $-$ &  $-$ &  \checkmark 
\\
\hline
\cellcolor{gray!25}$\calO_{22} = \left({i \bm{q}\over m_e}\times\pve\right)\cdot\pSxt\cdot\pSe + \pve\cdot\pSxt\cdot\left({i \bm{q}\over m_e}\times\pSe\right)$ &  \xmark & $\color{cyan} qv$ &  $-$ &  $-$ &  \checkmark 
\\
\hline
\cellcolor{gray!25}$\calO_{23} = - {i \bm{q}\over m_e}\cdot\pSxt\cdot(\pve \times \pSe)$ &  \xmark & $\color{cyan}  qv$ &  $-$ &  $-$ &  \checkmark 
\\
\hline
\cellcolor{gray!25}$\calO_{24} = {\bm{q} \over m_e} \cdot \pSxt \cdot  \left({ \bm{q} \over m_e} \times\pve\right)$ 
&  \xmark & $\color{orange} q^2 v$ &  $-$ &  $-$ &  \checkmark 
\\
\hline
\cellcolor{gray!25}$\calO_{25} = \left({\bm{q}\over m_e} \cdot \pSxt\cdot\pve \right) 
\left({\bm{q}\over m_e} \cdot\pSe \right)$ &  \xmark & $\color{orange} q^2 v$ &  $-$ &  $-$ &  \checkmark 
\\
\hline
\cellcolor{gray!25}$\calO_{26} = \left({\bm{q}\over m_e} \cdot \pSxt \cdot {\bm{q}\over m_e} \right) (\pve\cdot\pSe)$ 
& \xmark & $\color{orange}q^2 v$ &  $-$ &  $-$ &  \checkmark 
\\
\hline
\end{tabular} }
\caption{The basis of NR operators for DM-electron interactions up to second order in $\bm{q}$ and linear order in $\pve$. In the second column, the checkmark ``$\checkmark$" implies the corresponding NR operator is consistent with the one listed in \cite{Catena:2019gfa,Catena:2022fnk}, while `` \xmark" means the operator is not covered in these works. A bar in the 4th (5th) column means there is no spin-dependent (tensor spin) operator for scalar (fermion) DM. $\calO_2$ and $\calO_{16}$ being quadratic in $\pve$ are not shown. All operators are made dimensionless, and their coefficients $c_i$ are then also dimensionless.}
\label{tab:NRop}
\end{table}

Since the incoming DM and target electron are both non-relativistic with the corresponding velocities $v_{\rm DM}\sim 10^{-3}$ and $v_e \sim \alpha_{\rm em} \sim 10^{-2}$, it is convenient to describe the DM-bound-electron scattering within the NR EFT framework. 
In this section, we provide a basis of general NR operators for DM-electron interactions up to spin-one DM particle. We adopt a similar treatment as in constructing the NR operators of DM-nucleon interactions \cite{DelNobile:2021wmp}. The NR operators should preserve rotational and Galilean invariance, and the building blocks consist of kinematic variables and spin operators for each species. The former includes the DM momentum transfer $\bm{q}=\bm{p}-\bm{p}'$ with $\bm{p}$ ($\bm{p}'$) being the momentum of the initial (final) DM state, and the transverse DM-electron velocity in the NR limit, $\pve=\bm{v}-{\bm{q}/(2\mu_{xe}})-{\bm{k}/ m_e}$. Here $\bm{v}$ is the incoming DM velocity in the lab frame, $\bm{k}$ is the momentum of the initial electron, and $\mu_{xe}\equiv m_e m_x/(m_x+m_e)$ the reduced DM-electron mass where $m_e$ ($m_x$) is the electron (DM) mass. The $\bm{q}$ variable is usually accompanied by the imaginary unit $i$ to make it self-conjugate. 

For the electron (fermion or vector DM), we denote the identity and spin operators by $\mathds{1}_e$ and $\bm{S}_e$ ($\mathds{1}_x$ and $\bm{S}_x$), 
where the subscript $x=\phi, \chi, X$ stands for the scalar, fermion, or vector DM case, respectively. For the vector DM, besides $\mathds{1}_x$ and $\bm{S}_x$, a rank-two traceless spin tensor operator $\pmb{\tilde\calS}_x$ is needed \cite{Gondolo:2020wge}, 
\begin{align}
\tilde{\calS}_x^{ij} = 
{1\over2}\left(S_x^i S_x^j+S_x^j S_x^i \right)-{2\over3}\delta^{ij}, 
\label{eq:Sxij}
\end{align}
which is related to the symmetric spin operator $\bm{\calS}_x$ in \cite{Catena:2019hzw} by
$\tilde{\calS}_x^{ij}= \delta^{ij}/3 -\calS_x^{ij}$. Here the identity operators are, $\mathds{1}_\phi=1$, $\mathds{1}_{\chi,e}=\mathds{1}_{2\times2}$, and $\mathds{1}_X=\mathds{1}_{3\times3}$. The spin operators $\bm{S}_{\chi,e}$ and $\bm{S}_X$ are the usual traceless matrices in the normalization convention, 
\begin{align}
\tr[S_{\chi,e}^i S_{\chi,e}^j]={1\over2}\delta^{ij}, 
\quad 
\tr[S_X^i S_X^j]=2\delta^{ij},
\end{align}
and the rank-two tensor operator for vector DM is normalized using \cref{eq:Sxij} as follows, 
\begin{align}
\tr[  \tilde{\calS}_X^{mn} \tilde{\calS}_X^{ij}] = {1\over 2}(\delta^{im} \delta^{jn} + \delta^{in}\delta^{jm})
  -{1\over 3}\delta^{ij}\delta^{mn}.    
\end{align}
Note that the above choice automatically implies that the trace of the product of the rank-two tensor operator and the spin operator vanishes, i.e., $\tr[S_X^i \tilde{\calS}_X^{jk}]=0$. This means there will be no interference contribution to the scattering rate among the NR operators with $\mathds{1}_x$, $\bm{S}_x$, and $\pmb{\tilde\calS}_x$, and similarly for the operators with $\mathds{1}_e$ and $\bm{S}_e$ (see the results for DM response functions in \cref{tab:dmRF}).

Employing the aforementioned building blocks, $\{\mathbb{1}_e, \bm{S}_e\}\otimes\{\mathbb{1}_x, \bm{S}_x, \pmb{\tilde\calS}_x\}\otimes \{i \bm{q}, \pve\} $, one can construct the relevant NR operators for the DM-electron interactions. Most of the operators can be obtained from those available in the literature by relabeling the nucleon field by the electron field. For the scalar and fermion DM scenarios, the widely used NR operators are given in \cite{DelNobile:2018dfg,Fitzpatrick:2012ix}, while for the vector DM case the NR interactions induced in some simplified models are discussed in \cite{Catena:2019hzw,Dent:2015zpa}. 
With our notation, we list in \cref{tab:NRop} the most general LO non-redundant NR operators that are at most of second order in $\bm{q}$ and linear order in $\pve$. 
The power counting for each operator is provided in the third column. Note that only operators $\calO_{1,3,7,10}$ are present in the scalar DM case. The first fourteen operators from $\calO_1$ to $\calO_{15}$ are relevant to both the fermion and vector DM cases, which take the same convention as in \cite{Catena:2022fnk,Catena:2019gfa}. The remaining operators from $\calO_{17}$ to $\calO_{26}$ all involve the rank-two traceless spin operator $\pmb{\tilde\calS}_x$ and are specific to the vector DM case. The details for constructing these operators are provided in \cref{app:NRop}. The relations between the operators $\calO_{17}-\calO_{20}$ and those listed in \cite{Catena:2022fnk} are shown in the second column in \cref{tab:NRop}. It is worth noting that the operators $\calO_{21}$ and $\calO_{22}$ have no contributions to the nuclear recoil signal in the one-nucleon current approximation to the DM-nucleon scattering \cite{Gondolo:2020wge}. Our basis of complete and independent operators is universal and does not depend on any details of a fundamental theory of DM. 
In the next section we will consider the DM-electron interactions from the more fundamental relativistic EFT perspective and establish the connection between the NR EFT and relativistic EFT operators.

\section{Relativistic operators and their NR reduction}
\label{sec:NRreduct}

In this section we consider the DM-electron scattering induced by Lorentz invariant interactions. For light DM with a mass smaller than the electroweak scale, it is appealing to adopt the LEFT-like framework to describe the interactions between the DM and SM particles that satisfy the QCD and QED symmetries $\rm SU(3)_{\rm c}\times U(1)_{\rm em}$. With the merit of model independence, the interactions have been extensively studied within the LEFT-like framework, for instance, in the DM direct and indirect detections \cite{Brod:2017bsw,Baumgart:2022vwr,Aebischer:2022wnl}, and the meson and baryon decays involving light DM particles \cite{He:2022ljo,Lehmann:2020lcv,Li:2021phq}, etc. The LEFT-like framework can be extended to the SMEFT-like framework when we are interested in DM interactions with the whole spectrum of the standard model at a higher energy scale \cite{Liao:2016qyd,Criado:2021trs,Aebischer:2022wnl,Song:2023jqm}. 

The LEFT description of interactions between the scalar/fermion DM and SM light particles ($u,d,s,c,b$ quarks, all charged leptons and neutrinos, gluons, and photon) has been widely investigated in the literature \cite{Brod:2017bsw, Goodman:2010ku,Kumar:2013iva,Bishara:2017pfq,Badin:2010uh,Liang:2023yta}. 
In this work, we focus on both the contact DM-electron and the DM-photon interactions that induce the short-distance (SD) and the long-distance (LD) DM-bound-electron scattering, respectively. The effective operators up to dim 6 are listed in the second column of \cref{tab:NRmatch:sf}. For the scalar DM $\phi$, there are 4 DM-electron operators up to dim 6 and 2 additional DM-photon interactions characterizing the DM millicharge ($Q$) and the charge radius (cr). As for the fermion DM $\chi$, we have 10 dim-6 contact DM-electron operators and 5 DM electromagnetic (EM) operators (including the millicharge, magnetic and electric dipole moments, charge radius, and anapole moment). When $\phi$ is real or $\chi$ is Majorana, some of the operators vanish identically as indicated by the symbol ``$(\times)$". 

\begin{table}
\center
\resizebox{\linewidth}{!}{
\renewcommand\arraystretch{1.}
\begin{tabular}{| c | l | c |}
\hline
Dim & \multicolumn{1}{|c|}{ Relativistic operators} & NR reduction
\\
\hline
\multicolumn{3}{|c|}{\cellcolor{magenta!25}Scalar case}
\\
\hline
\multirow{2}{*}{dim-5} & $\calO^S_{\ell \phi}=(\overline\ell\ell) (\phi^\dagger\phi)$
& $2 m_e \calO_1$
\\
\cline{2-3}
 & $\calO^P_{\ell \phi}=(\overline\ell i \gamma_5 \ell) (\phi^\dagger\phi)$
& $-2 m_e \calO_{10}$
\\
\hline
\multirow{2}{*}{dim-6} & $\calO^V_{\ell \phi}=(\overline\ell \gamma^\mu \ell) (\phi^\dagger i\overleftrightarrow{\partial_\mu} \phi)\,(\times)$
& $ 4 m_e m_\phi  \calO_1$
\\
\cline{2-3}
 & $\calO^A_{\ell \phi}=(\overline\ell \gamma^\mu\gamma_5 \ell) (\phi^\dagger i\overleftrightarrow{\partial_\mu} \phi)\,(\times)$
& $ -8 m_e m_\phi  \calO_7$
\\
\Xhline{3\arrayrulewidth}
 & {\cellcolor{gray!25}$\calL_{\phi}^{Q}= 
 |(\partial_\mu  - i Q_\phi e A_\mu)\phi|^2
 \,(\times)$} 
 & $-4 Q_\phi e^2 { m_e m_\phi \over \bm{q}^2} \calO_1$ 
\\
\cline{2-3}
 & {\cellcolor{gray!25}$\calL_{\phi}^{\rm cr}= b_\phi (\phi^\dagger i\overleftrightarrow{\partial^\mu} \phi ) \partial^\nu F_{\mu\nu} \,(\times)$} &  $4 b_\phi e m_e m_\phi \calO_1$ 
\\
\hline
\hline
\multicolumn{3}{|c|}{\cellcolor{magenta!25}Fermion case}
\\
\hline
 \multirow{10}{*}{dim-6} & $\calO^S_{\ell \chi 1}=(\overline\ell\ell) (\overline\chi\chi)$
& $4  m_e m_\chi \calO_1$
\\
\cline{2-3}
 & $\calO^S_{\ell \chi 2}=(\overline\ell\ell) (\overline\chi i\gamma_5 \chi)$
& $4 m_e^2 \calO_{11}$
\\
\cline{2-3}
 & $\calO^P_{\ell \chi 1}=(\overline\ell i\gamma_5 \ell) (\overline\chi  \chi)$
& $-4 m_e m_\chi \calO_{10}$
\\
\cline{2-3}
 & $\calO^P_{\ell \chi 2}=(\overline\ell i\gamma_5 \ell) (\overline\chi i\gamma_5 \chi)$
&  $4 m_e^2 \calO_6$
\\
\cline{2-3}
 & $\calO^V_{\ell \chi 1}=(\overline\ell \gamma^\mu \ell) (\overline\chi \gamma_\mu \chi)\,(\times)$
& $4  m_e m_\chi \calO_1$
\\
\cline{2-3}
 & $\calO^V_{\ell \chi 2}=(\overline\ell \gamma^\mu \ell) (\overline\chi \gamma_\mu \gamma_5 \chi)$
& $8 m_e m_\chi (\calO_8-\calO_9)$
\\
\cline{2-3}
 & $\calO^A_{\ell \chi 1}=(\overline\ell \gamma^\mu\gamma_5 \ell) (\overline\chi \gamma_\mu \chi) \,(\times)$
&  $ -8 m_e  ( m_\chi \calO_7 + m_e\calO_9)$
\\
\cline{2-3}
 & $\calO^A_{\ell \chi 2}=(\overline\ell \gamma^\mu\gamma_5 \ell) (\overline\chi \gamma_\mu\gamma_5 \chi)$
& $-16 m_e m_\chi \calO_4$
\\
\cline{2-3}
 & $\calO^T_{\ell \chi 1}=(\overline\ell \sigma^{\mu\nu} \ell) (\overline\chi \sigma_{\mu\nu} \chi) \,(\times)$
& $32 m_e m_\chi \calO_4$
\\
\cline{2-3}
 & $\calO^T_{\ell \chi 2}=(\overline\ell \sigma^{\mu\nu} \ell) (\overline\chi i \sigma_{\mu\nu}\gamma_5 \chi) \,(\times)$
& $8 m_e(m_e\calO_{10}-m_\chi \calO_{11}-4m_\chi \calO_{12})$
\\
\Xhline{3\arrayrulewidth}
 & {\cellcolor{gray!25}$\calL_{\chi}^{Q}=
 \overline\chi i\gamma^\mu (\partial_\mu - i Q_\chi e A_\mu)\chi\,(\times)$} 
 & $-4 Q_\chi e^2\, { m_e m_\chi \over \bm{q}^2} \calO_1$
\\
\cline{2-3}
 & {\cellcolor{gray!25}${\cal L}_\chi^{\rm mdm}=
 \mu_\chi (\overline\chi\sigma^{\mu\nu}\chi) F_{\mu\nu} \,(\times)$}
& $4 \mu_\chi e\left( m_e \calO_1 +4 m_\chi \calO_4 
+{4 m_e^2 m_\chi\over \bm{q}^2}\left( \calO_5-\calO_6 \right) \right)$
\\
\cline{2-3}
 & {\cellcolor{gray!25}${\cal L}_\chi^{\rm edm}=d_\chi (\overline\chi i \sigma^{\mu\nu} \gamma_5 \chi) F_{\mu\nu} \,(\times)$}
& $ d _\chi e{16 m_e^2 m_\chi \over \bm{q}^2}\calO_{11 }$
\\
\cline{2-3}
 & {\cellcolor{gray!25}${\cal L}_\chi^{\rm cr}=b_\chi (\overline\chi \gamma^\mu\chi) \partial^\nu F_{\mu\nu} \,(\times)$}
& $4 b_\chi e \, m_e m_\chi \calO_1$
\\
\cline{2-3}
 & {\cellcolor{gray!25}${\cal L}_\chi^{\rm anap.}=a_\chi (\overline\chi \gamma^\mu\gamma_5 \chi) \partial^\nu F_{\mu\nu}$}
& $8 a_\chi e\, m_e m_\chi \left(\calO_8-\calO_9\right)$
\\
\hline
\end{tabular} 
}
\caption{Relativistic operators (left column) involving a scalar ($\phi$) or fermion ($\chi$) DM particle reduce in nonrelativistic limit to NR operators (right column) in \cref{tab:NRop}. The symbol ``$(\times)$" indicates vanishing of the operator for a real scalar ($\phi=\phi^\dagger$) or a Majorana fermion ($\chi^{\tt C}=\chi$). The EM Lagrangians highlighted in gray contribute to the LD DM-electron scattering via the exchange of a photon. 
} 
\label{tab:NRmatch:sf}
\end{table}

The LEFT operators involving a pair of vector DM and a pair of quark or lepton fields were systematically studied in \cite{He:2022ljo,Liang:2023yta}. Following \cite{He:2022ljo}, we consider two vector DM scenarios. In case A, the operators are constructed using the four-vector potential $X_\mu$. There are 4 independent dim-5 operators of the form $\overline\ell \ell X^\dagger X$ and 12 dim-6 operators with an additional covariant derivative in the form $\overline\ell \ell X^\dagger X D_\mu$. In case B, the DM field strength tensor $X_{\mu\nu}=\partial_\mu X_\nu-\partial_\nu X_\mu$ is used to construct the operators, and there are 6 dim-7 operators of the form $\overline\ell\ell X_{\mu\nu}^2$ at LO. All of these operators are shown in the left column of \cref{tab:NRmatch:v}. Similarly, the operators marked by a ``$(\times)$" vanish for the real vector DM case, i.e., $X_\mu^\dagger=X_\mu$. Notice that for the two operators $\calO_{\ell X1}^{\tt V}$ and $\calO_{\ell X1}^{\tt A}$, unlike \cite{He:2022ljo}, we use the traceless symmetric tensor current with $\gamma_{(\mu} i \overleftrightarrow{D_{\nu)} } \equiv {1\over 2}[ \gamma_{\mu} i \overleftrightarrow{D_{\nu} } + \mu \leftrightarrow \nu] - {1\over 4} g_{\mu\nu}i\overleftrightarrow{\slashed{D}}$.
For completeness, in \cref{tab:NRmatch:v}, we also list the EM operators of the vector DM up to dim 6 that are highlighted in gray.  
The EM operators for the vector DM together with their physical interpretation were considered previously in \cite{Hagiwara:1986vm,Gaemers:1978hg,Gounaris:1996rz,Hisano:2020qkq,Chu:2023zbo}.   
By using the integration by parts (IBP) and the Bianchi identity (BI), 
we find the two operators $\calO_{X\gamma 2}$ and $\calO_{X\gamma 5}$ were missed in these works.

\begin{table}
\center
\resizebox{\linewidth}{!}{
\renewcommand\arraystretch{1.2}
\begin{tabular}{| c | l | c |}
\hline
Dim & \multicolumn{1}{|c|}{ Relativistic operators} & NR reduction
\\
\hline
\multicolumn{3}{|c|}{\cellcolor{magenta!25}Vector case A}
\\
\hline
\multirow{4}{*}{dim-5} & $\calO_{\ell X}^{\tt S} 
= (\overline{\ell } \ell )(X_\mu^\dagger X^\mu) $ 
& $-2  m_e \calO_1 $  
\\
\cline{2-3}
 & $\calO_{\ell X}^{\tt P} 
=(\overline{\ell }i \gamma_5 \ell )(X_\mu^\dagger X^\mu)$ 
& $ 2  m_e \calO_{10}$ 
\\
\cline{2-3}
 & $\calO_{\ell X1}^{\tt T} 
= {i \over 2} (\overline{\ell }  \sigma^{\mu\nu} \ell ) (X_\mu^\dagger X_\nu - X_\nu^\dagger X_\mu),  \, (\times)$ 
& $-4 m_e \calO_4$ 
\\
\cline{2-3}
 & $\calO_{\ell X2}^{\tt T} 
= {1\over 2} (\overline{\ell }\sigma^{\mu\nu}\gamma_5 \ell ) (X_\mu^\dagger X_\nu - X_\nu^\dagger X_\mu),  \, (\times)$ 
& $ -m_e\left(\calO_{11}+4\calO_{12}\right)+4{m_e^2\over m_X }\left({1\over 3}\calO_{10}-\calO_{18}\right)$ 
\\
\Xhline{3\arrayrulewidth}
\multirow{12}{*}{dim-6} & $ \calO_{\ell X1}^{\tt V} 
= {1\over 2} [ \overline{\ell }\gamma_{(\mu} i \overleftrightarrow{D_{\nu)} } \ell ] (X^{\mu \dagger} X^\nu + X^{\nu \dagger} X^\mu  )$ 
& $ m_e^2 \calO_1$
\\
\cline{2-3}
 & $\calO_{\ell X2}^{\tt V} 
= (\overline{\ell }\gamma_\mu \ell )\partial_\nu (X^{\mu \dagger} X^\nu + X^{\nu \dagger} X^\mu)$
& $ -4 m_e^2\left(\calO_{17}+\calO_{20}\right)+{4\over 3} m_e (i\bm{q}\cdot\pve)\calO_1$ 
\\
\cline{2-3}
 & $\calO_{\ell X3}^{\tt V} 
= (\overline{\ell }\gamma_\mu \ell )( X_\rho^\dagger \overleftrightarrow{\partial_\nu} X_\sigma )\epsilon^{\mu\nu\rho\sigma}$ 
& $-4 m_e m_X \left(\calO_8-\calO_9\right)$ 
\\
\cline{2-3}
 & $\calO_{\ell X4}^{\tt V}
=(\overline{\ell}\gamma^\mu \ell )(X_\nu^\dagger i\overleftrightarrow{\partial_\mu} X^\nu), 
 \, (\times)$ & $-4 m_e m_X \calO_1$ 
\\
\cline{2-3}
& $\calO_{\ell X5}^{\tt V}
=(\overline{\ell }\gamma_\mu \ell)i\partial_\nu (X^{\mu \dagger} X^\nu - X^{\nu \dagger} X^\mu),\,(\times)$ 
& $2 m_e^2\left(\calO_5-\calO_6-{m_e\over m_X}\calO_{19}\right)+2\bm{q}^2\calO_4+{2\over 3}{{m_e}\over{m_X}}\bm{q}^2 \calO_1$ 
\\
\cline{2-3}
 & $\calO_{\ell X6}^{\tt V}
=(\overline{\ell }\gamma_\mu \ell) i\partial_\nu (X^\dagger_\rho X_\sigma )\epsilon^{\mu\nu\rho\sigma},
\, (\times) $ & $ -2 m_e^2 \calO_{11}$ 
\\
\cline{2-3}
  &  $\calO_{\ell X1}^{\tt A}
={1\over 2} [\overline{\ell }\gamma_{(\mu} \gamma_5 i \overleftrightarrow{D_{\nu)} }  \ell ](X^{\mu \dagger} X^\nu + X^{\nu \dagger} X^\mu )$ 
& $ -2  m_e^2\left({m_e\over m_X}\calO_9-4 \calO_{21}+{4\over 3}\calO_7\right) $ 
\\
\cline{2-3}
 & $\calO_{\ell X2}^{\tt A}
=(\overline{\ell }\gamma_\mu \gamma_5 \ell )\partial_\nu (X^{\mu \dagger} X^\nu + X^{\nu \dagger} X^\mu )$
& $-8 m_e^2\left({1\over3}\calO_{10}-\calO_{18}\right)$ 
\\
\cline{2-3}
 & $\calO_{\ell X3}^{\tt A}
=(\overline{\ell }\gamma_\mu\gamma_5 \ell ) (X_\rho^\dagger \overleftrightarrow{ \partial_\nu} X_\sigma )\epsilon^{\mu\nu\rho\sigma}$ & $8 m_e m_X \calO_4$ 
\\
\cline{2-3}
 & $\calO_{\ell X4}^{\tt A}
=(\overline{\ell }\gamma^\mu\gamma_5 \ell )(X_\nu^\dagger  i \overleftrightarrow{\partial_\mu} X^\nu)$ & $8 m_e m_X \calO_7$ 
\\
\cline{2-3}
 & $\calO_{\ell X5}^{\tt A} 
=(\overline{\ell }\gamma_\mu \gamma_5 \ell )i \partial_\nu (X^{\mu \dagger} X^\nu - X^{\nu \dagger} X^\mu  ),  \, (\times)$ & $ 4 m_e^2 \calO_9$ 
\\
\cline{2-3}
 & $\calO_{\ell X 6}^{\tt A}
=(\overline{\ell }\gamma_\mu\gamma_5\ell)i \partial_\nu (X^\dagger_\rho X_\sigma)\epsilon^{\mu\nu\rho\sigma},\,(\times)$ 
& $4 m_e^2 \left(\calO_{14}-{m_e\over m_X}\calO_{20}\right)$ 
\\
\Xhline{3\arrayrulewidth}
 & {\cellcolor{gray!25}$\calL_{\kappa_\Lambda}=i {{\kappa_\Lambda}\over 2}(X_\mu^\dagger X_\nu-X_\nu^\dagger X_\mu) F^{\mu\nu} \, (\times) $ } 
&  $ -2e \kappa_\Lambda \left[ {m_e \over {m_X}}
\left( {1\over 3}\calO_1 - {m_e^2 \over \bm{q}^2} \calO_{19}\right)
-\calO_4-{m_e^2 \over \bm{q}^2}\left(\calO_5-\calO_6\right)\right]$
\\
\cline{2-3}
 & {\cellcolor{gray!25}$\calL_{\tilde\kappa_\Lambda}= i 
 {\Tilde{\kappa}_\Lambda \over 2}(X_\mu^\dagger X_\nu-X_\nu^\dagger X_\mu) \tilde F^{\mu\nu} \, (\times) $} 
&  $2 e \Tilde{\kappa}_\Lambda  m_e^2{1\over \bm{q}^2} \calO_{11}$
\\
\hline
\multirow{5}{*}{dim-6} & {\cellcolor{gray!25}$\calO_{X\gamma 1}= \epsilon^{\mu\nu\rho\sigma} \left( X_\rho^\dagger \overleftrightarrow{\partial_\nu}X_\sigma\right)\partial^\lambda F_{\mu\lambda}$ } 
&  $-4 e m_e m_X \left(\calO_8-\calO_9\right)$
\\
\cline{2-3}
 & {\cellcolor{gray!25}$ \calO_{X\gamma 2}= \epsilon^{\mu\nu\rho\sigma} i \partial_\nu\left( X_\rho^\dagger X_\sigma\right)\partial^\lambda F_{\mu\lambda} \,(\times)$ } 
&  $-2 e m_e^2 \calO_{11}$
\\
\cline{2-3}
 & {\cellcolor{gray!25}$ \calO_{X\gamma 3}= \left( X_\nu^\dagger i\overleftrightarrow{\partial^\mu} X^\nu\right)\partial^\lambda F_{\mu\lambda}$ } 
&  $-4 e m_e m_X \calO_1$
\\
\cline{2-3}
& {\cellcolor{gray!25}$ \calO_{X\gamma 4}= \partial_\nu(X^{\mu\dagger} X^{\nu} + X^{\nu\dagger} X^{\mu}) \partial^\lambda F_{\mu\lambda} $ } 
&   $ 4e\, m_e\left[ {1\over 3} (i\bm{q}\cdot\pve)\calO_1 - m_e\left(\calO_{17}+\calO_{20}\right)\right] $ 
\\
\cline{2-3}
& {\cellcolor{gray!25}$ \calO_{X\gamma 5}=  i \partial_\nu(X^{\mu\dagger} X^{\nu} - X^{\nu\dagger} X^{\mu}) \partial^\lambda F_{\mu\lambda} \,(\times)$ } 
&  $e\left[2  m_e^2\left(\calO_5-\calO_6-{m_e\over m_X}\calO_{19}\right)+2\bm{q}^2\calO_4+{2\over 3}{{m_e}\over{m_X}}\bm{q}^2 \calO_1\right]$
\\
\hline
\multicolumn{3}{|c|}{\cellcolor{magenta!25}Vector case B}
\\
\hline
\multirow{6}{*}{dim-7} & $\tilde \calO_{\ell X1}^{\tt S} 
=(\overline{\ell }\ell )X_{\mu\nu}^\dagger  X^{\mu\nu}$ 
& $4 m_e m_X^2 \calO_1$ 
\\
\cline{2-3}
 & $\tilde \calO_{\ell X2}^{\tt S}
=(\overline{\ell }\ell )X_{\mu\nu}^\dagger \tilde X^{ \mu\nu}$
& $4 m_e^2 m_X \calO_{11}$ 
\\
\cline{2-3}
 & $\tilde \calO_{\ell X1}^{\tt P}
=(\overline{\ell }i \gamma_5\ell )X_{\mu\nu}^\dagger X^{ \mu\nu}$ 
& $-4 m_e m_X^2 \calO_{10}$
\\
\cline{2-3}
 & $\tilde \calO_{\ell X2}^{\tt P}
=(\overline{\ell }i \gamma_5\ell )X_{\mu\nu}^\dagger \tilde X^{ \mu\nu}$ 
& $4 m_e^2 m_X \calO_{6}$
\\
\cline{2-3}
 & $\tilde \calO_{\ell X1}^{\tt T}=
{i\over2} (\overline{\ell} \sigma^{\mu\nu} \ell )(X^{\dagger}_{ \mu\rho} X^{\rho}_{\,\nu}-X^{\dagger}_{\nu\rho} X^{\rho}_{\,\mu}),\,(\times) $ 
& $4 m_e m_X^2 \calO_{4}$
\\
\cline{2-3}
 & $\tilde \calO_{\ell X2}^{\tt T} = 
{1\over2}(\overline{\ell}\sigma^{\mu\nu}\gamma_5 \ell)(X^{\dagger}_{\mu\rho}X^{\rho}_{\,\nu}-X^{\dagger}_{\nu\rho} X^{\rho}_{\,\mu}),\,(\times)$ 
& ${1\over 3} m_e m_X \left[ 3 m_X( \calO_{11}+  4 \calO_{12} )
- 4 m_e (2 \calO_{10} + 3 \calO_{18} )
\right]$
\\
\Xhline{3\arrayrulewidth}
\multirow{3}{*}{dim-6} & {\cellcolor{gray!25}$ \Tilde{\calO}_{X\gamma 1}= i (X^{\dagger}_{ \mu\rho} X^{\rho}_{\,\nu}-X^{\dagger}_{ \nu\rho} X^{\rho}_{\,\mu}) F^{\mu\nu} \,(\times)$} & $ 2e \Big[{2\over 3}m_X (m_e+m_X)\calO_1 +2 m_X^2 \calO_4$
 \\
 & & $ +{1\over \bm{q}^2}\left( 2 m_e^2 m_X^2(\calO_5-\calO_6)-2 m_e^2 m_X(m_X-2 m
_e) \calO_{19}\right)  \Big]$
 \\
\cline{2-3}
  & {\cellcolor{gray!25}$ \Tilde{\calO}_{X\gamma 2}=i (X^{\dagger}_{ \mu\rho} X^{\rho}_{\,\nu}-X^{\dagger}_{ \nu\rho} X^{\rho}_{\,\mu}) \tilde F^{\mu\nu}\,(\times)$} & $ -4e m_e^2 m_X^2 {1\over\bm{q}^2} \calO_{11}$
 \\
 \hline
\end{tabular} }
\caption{Similar to \cref{tab:NRmatch:sf}, but for a vector DM $X$. The two dim-4 EM interactions are denoted in terms of Lagrangians. The double-arrow derivative is defined by $A\overset{\leftrightarrow}{\partial_\mu}  B = A (\partial_\mu B) - (\partial_\mu A) B $, and similarly for the covariant derivative $D_\mu$. } 
\label{tab:NRmatch:v}
\end{table}

To calculate the amplitude of DM-bound-electron scattering due to relativistic interactions given in \cref{tab:NRmatch:sf} and \cref{tab:NRmatch:v}, a natural way is through the matching onto the NR operators given in \cref{tab:NRop}. In such a way, each contribution to the scattering matrix element is factorized as a product of a free DM-electron matrix element and an atomic form factor, which will be discussed in the next section. This matching procedure is also known as the NR reduction, which involves the expansion of the relativistic amplitudes in the NR limit and then translates the obtained amplitudes as some NR operators. The NR reduction has been extensively applied in the DM-nucleon case to describe nuclear recoil signals \cite{DelNobile:2021wmp}. For scalar and fermion DM, the NR reduction for the most general Lorentz invariant DM-quark and DM-photon operators has been investigated in \cite{Bishara:2017pfq,DelNobile:2018dfg,DelNobile:2021wmp}.  
For the vector DM case, there is no systematic study on NR reduction of the effective operators given in \cref{tab:NRmatch:v} to our knowledge, except for a few operators generated in some simplified models \cite{Catena:2019hzw,Dent:2015zpa,Catena:2022fnk}. 
In the right column of \cref{tab:NRmatch:sf} and \cref{tab:NRmatch:v}, we show the NR matching results for each LEFT-like interaction considered in this work. For the scalar and fermion DM, our results are consistent with those given in \cite{Bishara:2017pfq,DelNobile:2018dfg}, while for the vector DM, we provide a more general and complete dictionary to the study of spin-one DM in direct detection experiments. In the rest part of this section, we outline the main procedures to reach these results. 

The NR reduction is conducted for the Lorentz invariant operators one by one. For each operator, we first write down its contribution to the Lorentz invariant amplitude $\calM$ for DM-free-electron scattering. For the three DM cases, the amplitudes have the general structures, 
\begin{subequations}
\begin{align}
\calM_{e-\phi} &\propto  \overline{u_e}(k')\Gamma  u_e(k), 
\\
\calM_{e-\chi} &\propto [\overline{u_e}(k')\Gamma_1 u_e(k)][\overline{u_\chi}(p')\Gamma_2 u_\chi(p)],
\\
\calM_{e-X} &\propto  [\overline{u_e}(k')\Gamma  u_e(k)] \epsilon_X^{s',\mu*}(p') \epsilon_X^{s,\nu}(p),
\end{align}
\end{subequations}
where $u_e(k)$ ($u_\chi(p)$) and $u_e(k')$ ($u_\chi(p')$) are respectively the spinor of incoming and outgoing electron (fermion DM), and $\epsilon_X^{s,\nu}(p)$ and $\epsilon_X^{s',\mu}(p')$ are respectively the polarization vector of the incoming and outgoing vector DM. Then we expand the contribution of each operator in the NR limit and truncate the result at its first nonvanishing order in $\pmb{q}$ or $\pve$. In the NR limit, the spinor is expanded in the chiral representation as 
\begin{align}
u^s (\pmb{p}) 
={1 \over \sqrt{4 m}} 
\begin{pmatrix}
(2 m-\pmb{p}\cdot\pmb{\sigma})\xi^s \\ (2 m+\pmb{p}\cdot\pmb{\sigma})\xi^s 
\end{pmatrix}+\calO(\pmb{p}^2),
\end{align}
where $\xi^s$ is the two-component Pauli spinor and $\sigma^i$ are the Pauli matrices. The polarization vector for a vector DM is 
\begin{align}
    \epsilon_X^{s,\mu}(\pmb{p})
    ={1\over m_X} 
\begin{pmatrix}
\pmb{p}\cdot\pmb{e}_X^s \\ m_X \pmb{e}_X^s
\end{pmatrix}+\calO(\pmb{p}^2),
\end{align}
where $\pmb{e}_X^s$ is the three-dimensional polarization vector. Now we translate the terms in amplitude to the building blocks for NR operators. In this manner, the kinematic variables are naturally changed to the corresponding operators, while the spinors and polarization vectors are substituted by the corresponding spin operators: 
\begin{subequations}
\begin{align}
& \xi^{s'\dagger}\xi^s \to \mathds{1}, \quad  
\xi^{s'\dagger} {\pmb{\sigma}\over 2} \xi^s \to \pmb{S},
\\
& \pmb{e}_X^{s'*}\cdot\pmb{e}_X^s\to \mathds{1}_X ,\quad 
(\pmb{e}_X^{s'*}\times \pmb{e}_X^s)\rightarrow i \pmb{S}_{X},
\\
& {1\over 3}(\pmb{e}_X^{s'*}\cdot\pmb{e}_X^s)\delta^{ij} 
- {1\over 2}\left(e_X^{s',i*} e_X^{s,j} + e_X^{s',j*} e_X^{s,i}\right)\rightarrow \tilde{\calS}_X^{ij},
\end{align}
\end{subequations}
where we have suppressed the indices for spin operators on the right hand side. The substitutions in the first line are applicable to both the electron and fermion DM. 

The first nonvanishing results of NR reduction due to each contact DM-electron and long-distance DM-photon interaction are provided in \cref{tab:NRmatch:sf} and \cref{tab:NRmatch:v}. Due to this strategy some NR operators with a higher power counting in $q$ or $v$ are not yet present for relativistic operators up to dim 6 or 7 that are considered here. For instance, the NR operators $\calO_{3,13,22,23}$ counted as $qv$ and all NR operators counted as $q^2 v$ are absent in this manner. The electromagnetic interactions could induce long-distance contributions, where a factor of $1/\bm{q}^2$ is attached to the NR operator. 
Finally, interference exists among different NR operators especially for the vector DM case.

\section{DM-atom scattering rate }
\label{sec:scatteringrate}

\subsection{Basic formalism for scattering rate}

For DM-electron scattering in an atomic target, we label the initial state of the bound electron by its atomic quantum numbers $(n,\ell,m)\equiv|1\rangle$ and the final state of the ionized electron by $(k^\prime,\ell^\prime,m^\prime)\equiv|2\rangle$ with $k'$ being its momentum. Here $n$ is the principal quantum number, $\ell (\ell')$ and $m (m')$ are the angular and magnetic quantum numbers relevant to the spherical harmonic function $Y_\ell^m (Y_{\ell'}^{m'})$, which is the angular part of the atomic wavefunction. Denoting the initial (final) three-momentum of DM by $\bm{p}\,(\bm{p}')$, the S-matrix element for the transition $|\bm{p}, 1\rangle\to |\bm{p}', 2\rangle $ takes the form \cite{Catena:2019gfa}, 
\begin{align}
\calS_{1\to 2} 
= - i \langle \bm{p}', 2| \int \td^4 x\,  {\cal H}_I(x)| \bm{p},1\rangle 
 \overset{{\rm NR}}{\Rightarrow} 2\pi \delta(E_f -E_i)  {V \over 4 m_x m_e}  
{\cal M}_{1\to 2},  
\label{eq:Smat}
\end{align}
where ${\cal H}_I(x)$ is the Hamiltonian density in the interaction picture derived from the NR operators in \cref{tab:NRop}
and $V$ is the normalization volume for single particle states. 
The total energies of the initial and final states are denoted by $E_i$ and $E_f$ respectively. In the NR limit, $E_i = m_x + m_x v^2/2 + m_e + E_1$ and $E_f =  m_x + |m_x \bm{v} - \bm{q}|^2 /(2m_x)+ m_e + E_2$, where $E_1~(E_2)$ is the binding (kinetic) energy of the initial (final) electron and $\bm{v}\,(v=|\bm{v}|)$ is the incoming DM velocity in the lab frame. The energy conservation condition implies that 
$\Delta E_{1\to 2}\equiv E_2-E_1 = \bm{q}\cdot \bm{v} - {q^2/(2m_x)}$ which serves as a constraint on $\bm{q}$ for given $E_{1,2}$. For the ionized electron, we will denote the kinetic energy $E_2$ by $E_e$, and $-E_1$ by $E_B^{n\ell}$ which is just the binding energy of the initial electron state. Thus, $\Delta E_{1\to 2} = E_e + E_B^{n\ell}$ in the practical calculation.
The second equality in \cref{eq:Smat} is obtained by going to the Schr\"odinger picture and inserting the momentum space completeness relation for the initial and final electron, with the transition amplitude given by 
\begin{align}
{\cal M}_{1\to 2} =\int {\td^3 \bm{k} \over (2 \pi)^3} \tilde\psi_{2}^*(\bm{k}+\bm{q}) 
 {\cal M}(\bm{q}, \pve) \tilde\psi_1(\bm{k}). 
 \label{eq:M1to2}
\end{align}
Here $\tilde \psi_1\equiv \tilde \psi_{n\ell m}$ and 
$\tilde \psi_2 \equiv \tilde \psi_{k'\ell' m'}$ are the initial and final electron wavefunctions in momentum space respectively. ${\cal M}(\bm{q}, \pve)$ is the matrix element for DM-free-electron scattering
that is only a function of the momentum transfer $\bm{q}\equiv \bm{p}-\bm{p}'$ and the DM-electron transverse velocity $\pve$. $\bm{k}$ is the momentum of the initial bound electron. With the matrix element in \cref{eq:M1to2}, the spin-averaged and -summed total rate of DM-atom scattering becomes \cite{Catena:2019gfa}
\begin{align}
\mathcal{R}_{1 \to  2} = {n_x \over 16 m_x^2 m_e^2} 
 \int {\td^3 \bm{q} \over (2 \pi)^3} 
 \int \td^3 \bm{v} \, f_x (\bm{v}) (2 \pi) \delta\left(E_f-E_i\right) \overline{\left|\calM_{1 \to 2}\right|^2},
 \label{eq:rate}
\end{align}
where $n_x$ is the local DM number density fixed by its energy density $\rho_x =n_x m_x = 0.4\, \rm GeV/cm^3 $ and $f_x(\bm{v})$ is the local DM three-dimensional velocity distribution, whose parameterization and numerical inputs are given below. 

To obtain the total ionization rate for a full atomic orbital ($n,\ell$), $\mathcal{R}_{\rm ion }^{n \ell}$, we have to sum
$\mathcal{R}_{1 \to 2}$  over the quantum numbers of the initial ($m$) and  final $(k',\ell',m')$ state electrons \cite{Essig:2015cda}, namely, 
\begin{align}
\mathcal{R}_{\rm ion }^{n \ell}
&= 2 \sum_{m=-\ell}^{\ell} 
\sum_{\ell^{\prime}=0}^{\infty} 
\sum_{m^{\prime}=-\ell^{\prime}}^{\ell^{\prime}} 
\int { V k^{\prime 2}\td k' \over  (2 \pi)^3} 
\mathcal{R}_{1 \to 2}
\nonumber
\\
&= \sum_{m=-\ell}^{\ell} \sum_{\ell^{\prime}=0}^{\infty} \sum_{m^{\prime}=-\ell^{\prime}}^{\ell^{\prime}} 
 \int \td \ln E_e {V k^{\prime 3} \over (2 \pi)^3} \mathcal{R}_{1 \to 2},
\label{eq:rionnl}
\end{align}
where the factor of 2 accounts for the electron's spin degeneracy and $E_e =  k'^2/(2m_e)$ is the kinetic energy of the ionized electron.
Together with \cref{eq:rate}, the differential rate becomes, 
\begin{align}
{\td \mathcal{R}_{{\rm ion}}^{n \ell} \over \td \ln E_e}=
{n_x \over 64 m_x^2 m_e^2} 
\int {\td^3 \bm{q} \over (2 \pi)^3} 
\int \td^3 \bm{v} f_x\, (\bm{v}) (2 \pi) \delta\left(E_f-E_i\right)
\overline{\left|\calM_{{\rm ion}}^{n \ell}\right|^2},  
\label{eq:dRdlnE}
\end{align}
where the kernel of integral 
\begin{align}
\overline{\left|\calM_{{\rm ion}}^{n \ell}\right|^2}  \equiv 
{4 V k^{\prime 3} \over (2 \pi)^3}
\sum_{m=-\ell}^{\ell} \sum_{\ell^{\prime}=0}^{\infty} 
\sum_{m^{\prime}=-\ell^{\prime}}^{\ell^{\prime}}  
\overline{\left|\calM_{1 \to 2}\right|^2}.
\label{eq:Mnlsqrd}
\end{align}

The calculation of $\overline{\left|\calM_{{\rm ion}}^{n \ell}\right|^2}$ is postponed to the next subsection, where one will find the dependence on $\bm{q}$ is through $\bm{q}^2$ and $\bm{q}\cdot \bm{v}$. We can then use the $\delta$-function to finish the angular integration of $\bm{q}$ in \cref{eq:dRdlnE}, giving 
\begin{align}
{\td \mathcal{R}_{{\rm ion}}^{n \ell} \over \td \ln E_e}=
{n_x \over 128 \pi m_x^2 m_e^2} 
\int \td q \, q \int {\td^3\bm{v} \over v} f_x (\bm{v}) 
\Theta (v-v_{\rm min}) 
\overline{\left|\calM_{{\rm ion}}^{n \ell}\right|^2}\Big|_{\theta_{qv}= \bar{\theta}_{q v}},
\end{align} 
where $\theta_{qv}$ is the angle between $\bm{q}$ and $\bm{v}$ with $\bar{\theta}_{q v}$ being the solution to the $\delta$-function ($ q v\cos\theta_{qv} = \Delta E_{1\to 2}+ q^2/(2m_x)$). $\Theta(v-v_{\rm min})$ is the Heaviside function with the minimum speed $v_{\rm min}= \Delta E_{1\to2}/q+q/(2m_x)$, which is achieved by requiring $\cos\bar \theta_{qv}=1$. Because the matrix element squared in the above integral depends on $\bm{v}$ only through its module $v$, the solid angle integration associated with $\bm{v}$ is completely encoded in the distribution function $f_x(\bm{v})$. Thus, we can define a one-dimensional DM velocity distribution, $F_x (v) \equiv v^2 \int d\Omega_x f_x(\bm{v} )$. After finishing the solid angle integration and changing the integration order of $q$ and $v$, we obtain, 
\begin{align}
{\td \mathcal{R}_{{\rm ion}}^{n \ell} \over \td \ln E_e}=
{n_x \over 128 \pi m_x^2 m_e^2} \int_{\sqrt{{2 \Delta E_{1\to 2}\over m_x}}}^{v_{\rm max}}  
{\td v\over v} F_x(v) \int_{q_-(v)}^{q_+(v)} \td q q   
\overline{\left|\calM_{{\rm ion}}^{n \ell}\right|^2} \Big|_{\theta_{qv} = \bar{\theta}_{q v}},
\label{eq:finalrate}
\end{align} 
where $q_{\pm}(v) = m_x v \pm\sqrt{m_x^2 v^2-2m_x \Delta E_{1\to 2}}$ and $v_{\rm max}$ is given below \cref{eq:vdis}. 

In the galactic rest frame, the DM velocity obeys a normal Maxwell-Boltzmann distribution
with the circular velocity $v_0= 220\,\rm km/s$ \cite{Smith:2006ym}, 
which leads to \cite{Lewin:1995rx,DelNobile:2021wmp}
\begin{align}
F_x (v) = 
{v \over \sqrt{\pi} v_0 v_\oplus}
\begin{cases}
 e^{-(v-v_\oplus)^2 / v_0^2}-e^{-(v+v_\oplus)^2 / v_0^2}, 
& \text{for}~ 0 \leq v \leq v_{\rm{esc}}-v_\oplus 
\vspace{0.1cm}
\\ 
 e^{-(v-v_\oplus)^2 / v_0^2}-e^{-v_{\rm{esc}}^2 / v_0^2}, 
& \text{for}~ v_{\rm{esc}}-v_\oplus<v \leq v_{\rm{esc}}+v_\oplus
\end{cases}.
\label{eq:vdis}
\end{align}
Here we adopt the averaged Earth relative velocity $v_\oplus = 244 \,\rm km/s$
and the escape velocity $v_{\rm esc} = 544\,\rm km/s$ \cite{Smith:2006ym}, leading to the maximal DM velocity in the lab frame,
$v_{\rm max}=v_{\rm esc}+v_\oplus=788\,\rm km/s$.

\subsection{Calculation of matrix element squared}

Now we calculate $\overline{\left|\calM_{{\rm ion}}^{n \ell}\right|^2}$ by starting from the DM-free-electron scattering matrix element $ {\cal M}(\bm{q}, \pve)$. 
Momentum conservation and Galilean invariance imply that only two out of the four three-dimensional momenta are actually independent. A convenient choice of independent variables is the momentum transfer $\bm{q}$ and the transverse velocity of the DM-electron system $\pve$. Note that $\delta\left(E_f-E_i\right)$ in \cref{eq:rate} is the energy conservation condition for DM-bound-electron scattering, and generally $\pve \cdot \bm{q} \neq 0$. Because all interested NR interactions are at most linear in $\pve$ (or equivalently, at most linear in $\bm{k}$), the amplitude for DM-free-electron scattering can be written as \cite{Liang:2024lkk} 
\begin{align}
\calM(\bm{q}, \pve) 
& =  
\calM(\bm{q}, 0)
+ \pve\cdot \nabla_{\pve} \calM(\bm{q}, \pve)
\nonumber
\\
&\equiv \calM_{\tt S} + \left( \bm{v}_{0}^\perp - { {\bm k}\over m_e} \right) \cdot {\bm \calM}_{\tt V},
\label{eq:Mfree}
\end{align}
where we have used the relation, $\pve = \bm{v}_0^\perp - { \bm{k}/ m_e}$,
and the following abbreviations 
\begin{align}
\calM_{\tt S} \equiv \calM(\bm{q}, 0), \quad 
{\bm \calM}_{\tt V} \equiv 
\nabla_{\pve} \calM(\bm{q}, \pve), \quad 
\bm{v}_0^\perp \equiv \bm{v} - { \bm{q}\over 2\mu_{xe}}.
\end{align}

Embedding the above free amplitude into the bound electron case in \cref{eq:M1to2} leads to the transition amplitude for DM-atom scattering as 
\begin{align}
\calM_{1\to 2} = f_{\tt S}(\bm{q}) \calM_{\tt S}  + {\bm f}_{\tt V}(\bm{q}) \cdot {\bm \calM}_{\tt V}.
\label{eq:M1to2new}
\end{align}
The two form factors $f_{\tt S}$ and $\bm{f}_{\tt V}$ are related to $f_{1\to2}(\bm{q})$ and $\bm{f}_{1\to2}(\bm{q})$ defined in \cite{Catena:2019gfa} by 
\begin{align}
f_{\tt S}(\bm{q}) = f_{1\to2}(\bm{q}), \quad 
\bm{f}_{\tt V}(\bm{q}) = \bm{v}_0^\perp f_{1\to2}(\bm{q}) - \bm{f}_{1\to2}(\bm{q}), 
\label{eq:ffrelation}
\end{align}
where 
\begin{subequations}
\label{eq:formfactor}
\begin{align}
f_{1 \to 2}(\bm{q})&= \int {\td^3 k \over (2 \pi)^3} \tilde\psi_2^*(\bm{k}+\bm{q}) \tilde\psi_1(\bm{k}), \\
\bm{f}_{1 \to 2}(\bm{q})&= \int {\td^3 k \over (2 \pi)^3} \tilde\psi_2^*(\bm{k}+\bm{q}){\bm{k} \over m_e} \tilde\psi_1(\bm{k}).    
\end{align}
\end{subequations}
Here $f_{1 \to 2}(\bm{q})$ is the usual scalar atomic form factor used widely in the literature and $\bm{f}_{1 \to 2}$ is a vectorial atomic 
form factor, which was first introduced in \cite{Catena:2019gfa}. From the NR operators in \cref{tab:NRop}, one can find the vector form factor is associated with those operators containing the transverse velocity $\pve$. In a recent paper \cite{Liang:2024lkk}, we found a minus sign mistake was made for the vector form factor in \cite{Catena:2019gfa} when Fourier transforming atomic wavefunctions from momentum to configuration space.
We leave our calculation details for these form factors in \cref{app:electronRF}.

\begin{table}
\center
\resizebox{\linewidth}{!}{
\renewcommand\arraystretch{1.4}
\begin{tabular}{| c | l |}
\hline
Type & \multicolumn{1}{|c|}{DM response functions} 
\\\hline
\multirow{5}*{\rotatebox[origin=c]{90}{\bf Scalar DM} } 
&  $a_0  = |c_1|^2+ {1\over 4} |c_{10}|^2 {\color{cyan} x_e}$ 
\\ 
&  $a_1  = {1\over 4} |c_7|^2 + {1\over 4} |c_3|^2 {\color{cyan}x_e}$
\\
& $a_2  = - {1\over 4} |c_3|^2 {\color{cyan}x_e}$ 
\\
& $a_3  ={1\over 2} {\Re}[c_3 c_7^*]$
\\
& $ a_4 =  -{1 \over 4} c_7 c_{10}^*$
\\\hline
\multirow{6}*{\rotatebox[origin=c]{90}{\bf Fermion DM} } 
& $a_0 = |c_1|^2+ {3 \over 16} |c_4|^2 +
\left( {1\over8}|c_9|^2+{1\over4} |c_{10}|^2+{1\over4} |c_{11}|^2 + {1\over8} {\Re}[c_4 c_6^*] \right) {\color{cyan} x_e }
+{1\over16} |c_6|^2 {\color{cyan} x_e^2 } $
\\
& $ a_1 =  {1\over4} |c_7|^2+{1\over4} |c_8|^2+{1\over8}|c_{12}|^2 
+\left( {1\over4} |c_3|^2+ {1\over4} |c_5|^2+ {1\over16} |c_{13}|^2+ {1\over16} |c_{14}|^2
-{1\over8} {\Re}[c_{12} c_{15}^*]\right) {\color{cyan}  x_e } $
\\
&$\quad\, +{1\over 16} |c_{15}|^2 {\color{cyan}  x_e^2} $
\\
& $ a_2  = - \left( 
{1\over4} |c_3|^2 + {1\over4}  |c_5|^2 
- {1\over8} {\Re}[c_{12} c_{15}^*]
- {1\over8}{\Re}[c_{13} c_{14}^*]
\right){\color{cyan}x_e} 
- {1\over 16} |c_{15}|^2 {\color{cyan}x_e^2} $
 \\
& $ a_3 = {1\over2} {\Re}[c_3 c_7^*] 
+ {1\over 2} {\Re}[c_5 c_8^*]
- {1\over 8} {\Re}[c_{12} c_{13}^*] 
+ {1\over8} {\Re}[c_{12} c_{14}^*] 
- {1\over 8} {\Re}[c_{14} c_{15}^*] {\color{cyan} x_e} $
 \\
& $ a_4 = {1\over16}c_4 c_{13}^* + {1\over16}c_4 c_{14}^* 
-{1\over4} c_7^* c_{10}-{1\over4} c_8^* c_{11}
-{1\over8} c_9 c_{12}^*
+\left({1\over16}c_6 c_{13}^*+{1\over16}c_6 c_{14}^*\right) {\color{cyan} x_e} $ 
\\\hline
\multirow{12}*{\rotatebox[origin=c]{90}{\bf Vector DM} } 
& $ a_0  =  |c_1|^2+{1\over 2} |c_4|^2
+\left( 
{1\over3} |c_9|^2 + {1\over4}|c_{10}|^2
+{2\over3}|c_{11}|^2 +{5\over36} |c_{18}|^2 
+{1\over3} {\Re}[c_4 c_6^*]
\right) {\color{cyan} x_e} $
\\
& $\quad\,+ \left(
{1\over6} |c_6|^2 
+{2\over9} |c_{19}|^2+{1\over12} |c_{20}|^2
\right) {\color{cyan}x_e^2 }$
\\%
& $  a_1  = {1\over4} |c_7|^2+{2\over3}|c_8|^2
+{1\over3} |c_{12}|^2 +{5\over36} |c_{21}|^2 
+ \left({1\over4} |c_3|^2+{2\over3} |c_5|^2
+{1\over6} |c_{13}|^2 + {1\over6} |c_{14}|^2
+ {1\over6} |c_{17}|^2 
 \right.$
\\
& $\quad\,\left. 
+ {3\over8} |c_{22}|^2
+{7\over72} |c_{23}|^2
-{1\over3} {\Re}[c_{12} c_{15}^*]
 +{1\over12} {\Re}[c_{21} c_{25}^*]
 -{1\over18} {\Re}[c_{21} c_{26}^*]
 +{1\over12} {\Re}[c_{22} c_{23}^*] \right) {\color{cyan} x_e} $
\\
& $\quad\,+\left({1\over6} |c_{15}|^2+{1\over6} |c_{24}|^2
+{1\over24} |c_{25}|^2+ {1\over18} |c_{26}|^2\right){\color{cyan} x_e^2} $
\\%
& $a_2  = 
- \left( {1\over 4} |c_3|^2+ {2\over3} |c_5|^2
-{1\over18} |c_{17}|^2+ {7\over 24}|c_{22}|^2
+ {1\over 72} |c_{23}|^2
-{1\over3}{\Re}[c_{12} c_{15}^*]
-{1\over3}{\Re}[c_{13} c_{14}^*]
 \right.$
\\
& $ \quad\, \left. 
- {1\over36}{\Re}[c_{21} c_{25}^*]
-{1\over 6} {\Re}[c_{21} c_{26}^*]
+{1\over4} {\Re}[c_{22} c_{23}^*] \right) {\color{cyan} x_e}  $
\\
& $\quad\,
- \left( {1\over 6}  |c_{15}|^2+{1\over 6}  |c_{24}|^2
- {1\over 72}  |c_{25}|^2 
- {1\over 9} {\Re}[c_{25} c_{26}^*] \right) {\color{cyan} x_e^2}$
\\%
& $ a_3  = 
{1\over 2} {\Re}[c_3 c_7^*]+ {4\over 3} {\Re}[c_5 c_8^*]
- {1\over 3} {\Re}[c_{12} c_{13}^*] +{1\over 3} {\Re}[c_{12} c_{14}^*] 
+{5\over 12} {\Re}[c_{21} c_{22}^*] +{5 \over 36} {\Re}[c_{21} c_{23}^*] $
\\
& $\quad\, - \left({ 1 \over 3} {\Re}[c_{14} c_{15}^*]
+ { 1 \over 3} {\Re}[c_{17} c_{24}^*] - { 1\over12} {\Re}[c_{22} c_{25}^*] 
- { 1\over12} {\Re}[c_{23} c_{25}^*]
+ { 1\over9} {\Re}[c_{23} c_{26}^*] \right) {\color{cyan} x_e}$
 \\%
 & $ a_4  = 
{1\over 6} c_4 c_{13}^*+{1\over 6} c_4 c_{14}^* 
- {1\over 4} c_7^* c_{10} - {2\over 3} c_8^* c_{11} 
- {1\over 3} c_9 c_{12}^*
- {5\over 36} c_{18} c_{21}^*$
\\
& $\quad\, + \left(
{1\over 6} c_6 c_{13}^* + {1\over 6}  c_6 c_{14}^*
+ {2\over 9}  c_{17}^* c_{19}-{1\over 12}  c_{20} c_{22}^*
+ {1\over 12} c_{20} c_{23}^* - {1\over 18} c_{18} c_{25}^*
-{1\over 18} c_{18} c_{26}^*\right) {\color{cyan} x_e }$
\\%
\hline 
\end{tabular} }
\caption{The general DM response functions for DM up to spin one. We highlight the dependence on $x_e=\bm{q}^2/m_e^2$ in cyan to help assess the relative importance of each term.
}
\label{tab:dmRF}
\end{table}

From \cref{eq:Mfree}, one can directly read off the matrix elements ${\cal M}_{\tt S}$ and $\bm{{\cal M}}_{\tt V}$ from the NR operators in \cref{tab:NRop} together with their corresponding WCs $c_i$. Upon taking the complex conjugate and squaring, we find the spin-averaged matrix element squared takes a compact form, 
\begin{align}
 \overline{|\calM_{1\to2}|^2} 
& = a_0 |f_{\tt S}|^2
+ a_1 |\bm{f}_{\tt V}|^2
+ {a_2 \over x_e}  \left|{\bm{q}\over m_e} \cdot \bm{f}_{\tt V} \right|^2
+ i\, a_3 {\bm{q} \over m_e}\cdot(\bm{f}_{\tt V}\times \bm{f}_{\tt V}^*)
+ 2\,\Im\left[a_4 f_{\tt S} \bm{f}_{\tt V}^* \cdot {\bm{q}\over m_e} \right].
\label{eq:M1to2sq}
\end{align}
The details of the calculation for the above result are given in \cref{app:dmRF}. The coefficients $a_{0,1,2,3,4}$ are the so-called DM response functions which depend only on the WCs of NR operators and $x_e={\bm{q}^2/m_e^2}$ and are summarized in \cref{tab:dmRF} for the three (scalar, fermion, vector) DM scenarios.
The WCs of NR operators are generally  complex to cover scenarios like inelastic DM, and the elastic DM-electron scattering is recovered by sending the WCs to be real. From \cref{tab:dmRF}, one can see that the square of each single WC appears in the first three terms $a_{0,1,2}$, while $a_{3,4}$ involve only the interference terms of different WCs. In the following, we will see that the $a_{3,4}$ terms for real WCs have no contribution to the elastic DM-atom scattering. 

To further simplify the results and make comparisons with the results in \cite{Catena:2019gfa}, we use the relations for the form factors in \cref{eq:ffrelation},
\begin{subequations}
\begin{align}
i {\bm{q} \over m_e}\cdot(\bm{f}_{\tt V}\times \bm{f}_{\tt V}^*)  
& = 
2 {\Im}\left[ f_{1\to 2} \bm{f}_{1\to 2}^* \cdot { (\bm{q}\times \bm{v}_0^\perp)\over m_e } \right]  
+ i {\bm{q} \over m_e}\cdot(\bm{f}_{1\to 2}\times \bm{f}_{1\to 2}^*),
\label{eq:a3coeff}
\\
f_{\tt S} \bm{f}_{\tt V}^* \cdot {\bm{q}\over m_e} 
& = 
y_e |f_{1\to 2}|^2 - f_{1\to 2} \bm{f}_{1\to 2}^* \cdot {\bm{q}\over m_e}, 
\label{eq:a4coeff}
\end{align}
\end{subequations}
where $ y_e \equiv  { \bm{q}\cdot \bm{v}_0^\perp/ m_e }$. 
Together with \cref{eq:ffrelation}, we obtain
\begin{align}
\overline{|\calM_{1\to2}|^2} 
 = &\left( a_0 + a_1 |\bm{v}_0^\perp|^2 + a_2 {y_e^2\over x_e} + 2 y_e {\Im}[a_4]\right) |f_{1\to 2}|^2
+ a_1 \left|\bm{f}_{1\to 2}\right|^2
+ { a_2 \over x_e}\left|\bm{f}_{1\to 2} \cdot {\bm{q} \over m_e} \right|^2
\nonumber
\\
& - 2 a_1 {\Re}[f_{1\to 2} \bm{f}_{1\to 2}^*\cdot \bm{v}_0^\perp ]
 - 2 \left( a_2 {y_e \over x_e} + {\Im}[a_4] \right) {\Re}\left[f_{1\to 2} \bm{f}_{1\to 2}^* \cdot {\bm{q}\over m_e} \right]
 \nonumber
 \\
&  + 2 a_3 \,{\Im}\left[f_{1\to 2} \bm{f}_{1\to 2}^* \cdot { (\bm{q}\times \bm{v}_0^\perp)\over m_e } \right]
+ a_3 \left[i {\bm{q} \over m_e}\cdot(\bm{f}_{1\to2}\times \bm{f}_{1\to2}^*) \right]
\nonumber
\\
&  - 2 {\Re}[a_4] {\Im}\left[f_{1\to 2} \bm{f}_{1\to 2}^* \cdot {\bm{q}\over m_e} \right].
\label{eq:M1to2sqnew}
\end{align}
Once summing over the magnetic quantum numbers $(m,m')$ of the initial and final atomic states, one finds the form factor combination 
$\sum_{m',m}f_{\tt 1\to2}\bm{f}_{1\to2}^*$ is real and proportional to $\bm{q}$, and $\sum_{m',m}\bm{q}\cdot(\bm{f}_{1\to2} \times \bm{f}_{1\to2}^*) =0$. The proofs for these results are provided in \cref{app:electronRF}. Thus the two terms associated with $a_3$ will have no contribution to the DM-atom scattering. Furthermore, as we will focus on the Hermitian NR operators with real WCs, which apply to a wide class of DM scenarios, $a_4$ is also real. Thus, the $a_4$ terms vanish as well upon summing over magnetic quantum numbers. Therefore, we will drop these terms in the following consideration. By the proportionality of $\sum_{m',m}f_{1\to2}\bm{f}_{1\to2}^* \propto {\bm q}$, we have 
\begin{align}
\sum_{m,m'} f_{1\to2}^* \bm{f}_{1\to2}\cdot  \bm{v}_0^\perp 
= {y_e \over x_e} \sum_{m,m'} f_{1\to2}^* \bm{f}_{1\to2}\cdot {\bm{q} \over m_e}. 
\end{align}
Consequently, the squared matrix element becomes, 
\begin{align}
\sum_{m',m}  \overline{|\calM_{1\to2}|^2}
& = a_0 I_0 + a_1 I_1 + a_2 I_2, 
\end{align}
where 
\begin{subequations}
\begin{align}
I_0 & \equiv  \sum_{m',m}   |f_{\tt S}|^2 =  \sum_{m',m} |f_{1\to 2}|^2,
\\
I_1 & \equiv  \sum_{m',m}  |\bm{f}_{\tt V}|^2 = \sum_{m',m} \left(|\bm{v}_0^\perp|^2 |f_{1\to 2}|^2  +  |\bm{f}_{1\to 2}|^2 - 2 {y_e \over x_e} \Re\left[ {f_{1\to 2}^* \bm{f}_{1\to 2} \cdot \bm{q}\over m_e} \right] \right),
\\
I_2 & \equiv \sum_{m',m}   {1\over x_e}\left|{\bm{q}\over m_e} \cdot \bm{f}_{\tt V} \right|^2 =  \sum_{m',m} \left(  { y_e^2\over x_e} |f_{1\to 2}|^2 
+ { 1\over x_e}\left|{\bm{f}_{1\to 2} \cdot\bm{q} \over m_e}\right|^2  
- 2 { y_e\over x_e} \Re\left[ {f_{1\to 2}^* \bm{f}_{1\to 2} \cdot \bm{q}\over m_e} \right] \right).
\end{align}
\end{subequations}
Taking the above expressions into \cref{eq:Mnlsqrd}, we finally get \cite{Liang:2024lkk}
\begin{align}
\overline{\left|\calM_{\rm{ion}}^{n \ell}\right|^2}=
a_0 \widetilde W_0 
+ a_1 \widetilde W_1 
+ a_2 \widetilde W_2.
\end{align}
The $\widetilde W_{0,1,2}$ are termed generalized atomic response functions, and are related to the four ones  $W_{1,2,3,4}$ defined in \cite{Catena:2019gfa} as follows \cite{Liang:2024lkk}, 
\begin{align}
\widetilde W_0 = W_1, \quad 
\widetilde W_1 = |\bm{v}_0^\perp|^2 W_1 -2 {y_e \over x_e} W_2 + W_3, \quad 
\widetilde W_2= {y_e^2 \over x_e} W_1  -2  {y_e\over x_e} W_2 + {1\over x_e}W_4,
\label{eq:W012tilde}
\end{align}
where
\begin{subequations}
\label{eq:W1to4}
\begin{align}
W_1 & = V {4 k'^3 \over {(2 \pi)^3}} \sum_{m=-\ell}^{\ell} \sum_{\ell'=0}^\infty \sum_{m'=-\ell'}^{\ell'} |f_{1\to 2}(\bm{q})|^2,
\\
W_2 & = V {4 k'^3 \over {(2 \pi)^3}} \sum_{m=-\ell}^{\ell} \sum_{\ell'=0}^\infty \sum_{m'=-\ell'}^{\ell'}  {\bm{q} \over m_e} \cdot f_{1\to 2}(\bm{q}) \bm{f}_{1\to2}^*(\bm{q}),
\\
W_3 & = V {4 k'^3 \over {(2 \pi)^3}} \sum_{m=-\ell}^{\ell} \sum_{\ell'=0}^\infty \sum_{m'=-\ell'}^{\ell'} |\bm{f}_{1 \to 2}(\bm{q})|^2,
\\
W_4 & = V {4 k'^3 \over {(2 \pi)^3}} \sum_{m=-\ell}^{\ell} \sum_{\ell'=0}^\infty \sum_{m'=-\ell'}^{\ell'} |{\bm{q} \over {m_e}} \cdot \bm{f}_{1\to 2}(\bm{q})|^2.
\end{align}
\end{subequations}
The atomic response functions depend on the orbital $(n,\ell)$ and momenta $(\bm{q},k')$, which are not shown for brevity, i.e., $W_{1,2,3,4}\equiv W_{1,2,3,4}^{n\ell}(k^\prime,\bm{q})$ and $\widetilde W_{0,1,2}\equiv \widetilde W_{0,1,2}^{n\ell}(k^\prime,\bm{q})$.
The calculation of $W_i$ is reviewed in \cref{app:electronRF}, and the parameters $|\bm{v}_0^\perp|^2$ and $y_e$ can be expressed as 
\begin{align}
|\bm{v}_0^\perp|^2
= v^2+ {x_e \over 4}\left( 1 - {m_e^2\over m_x^2}\right)
-{\Delta E_{1 \to 2} \over m_e} \left( 1 + {m_e\over m_x}\right)
, \quad 
y_e 
=  {\Delta E_{1 \to 2} \over m_e} - {x_e \over 2}.
\label{eq:v02andye}
\end{align}
Together with \cref{eq:finalrate}, one can readily compute the rate and set a sensitivity bound using experimental data. One should note that, for the practical case of a liquid detector the sensitivity range of the DM mass is  $m_x \gtrsim 5\,{\rm MeV}\approx 10\, m_e$, and 
$|\bm{v}_0^\perp|^2 \approx x_e/4-\Delta E_{1\to 2}/m_e\approx y_e^2/x_e$ to very good precision. Here the approximation in the second step comes from the fact that the $(\Delta E_{1\to 2})^2/m_e^2$ term in $y_e^2$ is negligible; e.g., it causes a correction below $2\,\%$ for typical values of $\Delta E_{1\to 2}\sim 60\,\rm eV$ and $|\bm{q}|\,\sim 20\,\rm keV$. Then we find
\begin{align}
\widetilde W_1 \approx {y_e^2 \over x_e} W_1 -2 {y_e \over x_e} W_2 + W_3.
\label{eq:Wtildeapp}
\end{align}
Note that this approximate $\widetilde W_1$, together with the $\widetilde W_{0,2}$, is independent of the DM parameters. Numerically, the correction due to the DM mass and the square term of $\Delta E_{1\to2}$ is sub-percent and thus can be totally neglected. In \cref{app:electronRF}, we will see this approximation is very helpful to calculate these atomic response functions, see \cref{eq:W1t,eq:W2t}. 

Now we make a comparison with the formalism in the literature. First, in \cite{Catena:2019gfa} there are four DM response functions $R^{n\ell}_{1,2,3,4}$ accompanying the four atomic response functions $W_{1,2,3,4}$. The functions $R^{n\ell}_{1,2,3,4}$ depend manifestly on the atomic orbital $(n,\ell)$ and are essentially the coefficients of the quadratic form factors in \cref{eq:M1to2sqnew}. The atomic dependence in $R^{n\ell}_{1,2,3,4}$ can be significant; for instance, in the coefficient of $|f_{1\to 2}|^2$ the atomic energy gap $\Delta E_{1\to 2}$ in the $y_e^2$ term of $a_2$ can be as large as $40\,\%$.
In contrast, our DM response functions $a_{0,1,2}$ are derived from the decomposition of scattering amplitude in terms of $\pve$ instead of $\bm{k}$, which only depend on the WCs of NR operators and $x_e$ without any atomic effect incurred. Our results imply there are close relationships among $R^{n\ell}_i$ leading to three compact combinations $a_{0,1,2}$: while $a_0$ captures only the velocity independent operators, $a_{1,2}$ are sensitive only to the velocity dependent ones. With the organization of DM response functions $a_i$ whose $x_e$ dependence is shown in \cref{tab:dmRF}, the leading order contributions of NR operators are clearly identified. Second, while the four atomic response functions $W_{1,2,3,4}$ in \cref{eq:W1to4} are independent of the DM properties, our three combinations $\widetilde W_{0,1,2}$ are also independent of DM for a liquid detector to very good precision (better than one percent). The main advantage is that $\widetilde W_{0,1,2}$ are more amenable to understand the cancellation mechanism and relative importance among $W_i$ as we have shown in \cite{Liang:2024lkk}. 
Furthermore, the expressions for $\widetilde W_{0,1,2}$ are actually very compact after taking into account the atomic wave functions, as can be seen in \cref{eq:W1t,eq:W2t}, and thus more efficient for numerical calculations.

\begin{table}
\center
\resizebox{\linewidth}{!}{
\renewcommand\arraystretch{1.7}
\begin{tabular}{|c|c|c|c|c|c|}
\hline
Class  & Chosen one & Scalar DM ($c_i^{\tt s}$) & Fermion DM ($c_i^{\tt f}$) & Vector DM ($c_i^{\tt v}$) & RFs
\\\hline%
\multirow{3}*{$a_0$} & $x_e^0: c_1^{\tt v}$ 
& $c_1^{\tt s} = c_1^{\tt v}$
& $c_{1,4}^{\tt f} =\left(1,\sqrt{16\over 3} \right) c_1^{\tt v}$
& $c_4^{\tt v}=\sqrt{2}\, c_1^{\tt v}$
& $\widetilde W_0$
\\\cline{2-6}%
& $x_e^1: c_{10}^{\tt v}$  
& $c_{10}^{\tt s} = c_{10}^{\tt v}$
& $c_{9,10,11}^{\tt f}= (\sqrt{2},1,1) c_{10}^{\tt v}$
& $c_{9,11,18}^{\tt v}= \left(\sqrt{3\over 4},\sqrt{3\over8}, \sqrt{9\over5}\right) c_{10}^{\tt v}$
& ${1\over 4} \widetilde W_0$
\\\cline{2-6}%
& $x_e^2: c_6^{\tt v}$  & $-$ 
&  $c_6^{\tt f} = \sqrt{8\over 3} c_6^{\tt v} $
& $c_{19,20}^{\tt v} = \left( \sqrt{3\over 4}, \sqrt{2} \right) c_6^{\tt v}$
& ${1\over 6} \widetilde W_0$
\\\hline
\multirow{9}*{$a_{1,2}$}  & $x_e^0: c_7^{\tt v}$ 
& $c_7^{\tt s} = c_7^{\tt v}$   
& $c_{7,8,12}^{\tt f} =(1,1,\sqrt{2}) c_7^{\tt v}$
& $c_{8,12,21}^{\tt v} = \left(\sqrt{3\over 8},\sqrt{3\over 4},\sqrt{9\over 5}\right) c_7^{\tt v}$
& ${1\over 4} \widetilde W_1$
\\\cline{2-6}%
& $x_e^1: c_3^{\tt v}$ & $c_3^{\tt s} = c_3^{\tt v}$ 
& $c_{3,5}^{\tt f} = (1,1) c_3^{\tt v}$
& $c_{5}^{\tt v} = \sqrt{3\over 8}\, c_3^{\tt v}$
& ${1\over 4}( \widetilde W_1-\widetilde W_2)$
\\\cline{2-6}%
& $x_e^1: c_{13}^{\tt v}$ & $-$  
& $c_{13,14}^{\tt f} = \left( \sqrt{8\over 3},\sqrt{8\over 3}\right) c_{13}^{\tt v}$
& $c_{14}^{\tt v} = c_{13}^{\tt v}$
& ${1\over 6}\widetilde W_1 $
\\\cline{2-6}%
& $x_e^2: c_{15}^{\tt v}$  & $-$ 
& $c_{15}^{\tt f} = \sqrt{8\over 3} c_{15}^{\tt v}$
& $c_{24}^{\tt v} = c_{15}^{\tt v}$
& ${1\over 6}( \widetilde W_1-\widetilde W_2)$
\\\cline{2-6}
&\cellcolor{gray!25} $x_e^1: c_{17}^{\tt v}$ & $-$ & $-$ & \checkmark 
& ${1\over 18}( 3 \widetilde W_1 + \widetilde W_2)$
\\\cline{2-6}
&\cellcolor{gray!25} $x_e^1: c_{22}^{\tt v}$ & $-$ & $-$ & \checkmark 
& ${1\over 24}( 9 \widetilde W_1 - 7 \widetilde W_2)$
\\\cline{2-6}
&\cellcolor{gray!25} $x_e^1: c_{23}^{\tt v}$ & $-$ & $-$ & \checkmark
& ${1\over 72}( 7 \widetilde W_1 - \widetilde W_2)$
\\\cline{2-6}
&\cellcolor{gray!25} $x_e^2: c_{25}^{\tt v}$ & $-$ & $-$ & \checkmark
& ${1\over 72}( 3 \widetilde W_1 + \widetilde W_2)$
\\\cline{2-6}
&\cellcolor{gray!25} $x_e^2: c_{26}^{\tt v}$ & $-$ & $-$ & \checkmark
& ${1\over 18}\widetilde  W_1$
\\\hline
\end{tabular} }
\caption{Upper bounds on WCs in \cref{eq:bmWC} for vector DM are converted to those on other WCs. 
In the 2nd column, $x_e^n:~c_i^{\tt v}$ means that $c_i^{\tt v}$ appears at $O(x_e^n)$ in $a_j$. In the 3rd (4th, 5th) column, an upper bound on a benchmark WC for vector DM is converted to those on various WCs for scalar (fermion, vector) DM using the displayed scaling factors read off the equations in \cref{tab:dmRF}. The checkmark in the 5th column means that the chosen benchmark WC stands alone and is not used for conversion. The last column shows the combination of generalized atomic RFs that accompanies $c^{\tt v}_i$ in the 2nd column.
}
\label{tab:rescal}
\end{table}

With limited data available, in the following numerical analysis, we will take one operator a time to set a sensitivity bound while neglecting their interference effect. For the case of NR operators, we use the vector DM case as a benchmark and choose the WCs: 
\begin{align}
\{c_1, c_3, c_6, c_7, c_{10}, c_{13}, c_{15}, c_{17}, c_{22}, c_{23}, c_{25}, c_{26}\}
\label{eq:bmWC}
\end{align}
to set a bound, which can be converted into upper bounds on the remaining ones by using the equations shown in \cref{tab:dmRF}. For convenience we display in \cref{tab:rescal} the rescaling factors for the conversion of upper bounds. 
In the second column of the table, we list these chosen WCs including their accompanying $x_e$ power whose corresponding atomic response functions are shown in the last column. The gray sector applies only to the vector DM case. 

\section{Constraints from xenon experiments}
\label{sec:expcon}

The process of a bound-state electron being ionized by DM-electron scattering is well described in \cref{sec:scatteringrate}. 
This primary ionized electron will further interact with surrounding atoms, ultimately converting its energy into ionized electrons, scintillation photons, and heat.
Owing to the delicate design of dual-phase time projection chambers, xenon experiments such as XENON1T, LUX-ZEPLIN (LZ), and PandaX-4T can effectively measure and distinguish between the scintillation photon signal (S1) and the ionized electron signal (S2).
Compared to the S1 signal, the S2 signal has a lower energy threshold, leading to a better capability for probing light DM. 
Therefore, in this work, we use the S2-only data from xenon experiments, including XENON10 \cite{XENON10:2011prx}, XENON1T \cite{XENON:2019gfn}, and PandaX-4T \cite{PandaX:2022xqx}, to constrain DM-electron EFT operators for DM masses below 10 GeV.
The LZ is not included due to the lack of S2-only  analysis.

\subsection{Numerical calculation}

We utilize the constant-$W$ model \cite{Essig:2012yx,Essig:2017kqs} to convert the ${\rm d} {\cal R}_{{\rm ion}}^{n \ell} / {\rm d} E_e$ spectrum into that of the number of ionized electrons, ${\rm d} {\cal R}_{{\rm ion}}^{n \ell} / {\rm d} n_e$.
The primary ionized electron with an energy $E_e$ can generate $n^{(1)}= \lfloor E_e/W\rfloor$ quanta, where $W=13.8\,\rm eV$ is the average energy required to create a single quantum \cite{Shutt:2006ed,Essig:2012yx}
and $\lfloor x\rfloor$ is the floor function to give the largest integer less than or equal to $x$.
If the primary electron is not the outermost shell electron, a photon can be emitted by de-excitation of the unstable xenon ion state, generating $n^{(2)} = \lfloor (E_B^{n\ell}-E_B^{\rm 5p})/W\rfloor$ additional quanta for the xenon target.
These $n^{(1)}+n^{(2)}$  quanta in total will become $N_i$ ions (${\rm Xe}^+$) and $N_{\rm ex}$ excited atoms (${\rm Xe}^*$) with $N_{\rm ex} / N_i \simeq 0.2$. 
A fraction of the ions can recombine with the ionized electrons to form other ${\rm Xe}^*$ with the probability of $f_R$.
Hence, the probability of an induced quantum becoming a detected ionized electron is given by $f_e = (1-f_R)(1+N_{\rm ex}/N_i)^{-1}$ and the probability of the primary ionized electron becoming a detected ionized electron is $1-f_R$.
Thus, with a primary ionized electron of energy $E_e$, the probability to detect $n_e$ ionized electron for the atomic orbital $(n, \ell)$ can be written as \cite{Essig:2012yx}
\begin{align}
P_{\rm ion}^{n\ell}(n_e|E_e) = (1-f_R)\, P_{\rm bin} (n_e -1, n^{(1)}+n^{(2)}, f_e) 
+ f_R \, P_{\rm bin} (n_e, n^{(1)}+n^{(2)}, f_e),
\end{align}
where $P_{\rm bin}$ is the binomial distribution with 
\begin{align}
P_{\rm bin} (n_{\rm suc}, n_{\rm try} , f_{\rm suc})
= {n_{\rm try}! \over n_{\rm suc}!(n_{\rm try}-n_{\rm suc})!} (f_{\rm suc})^{n_{\rm suc}} 
(1-f_{\rm suc})^{n_{\rm try}-n_{\rm suc}}.
\end{align}
In the above, $n_{\rm suc}$, $n_{\rm try}$, and $f_{\rm suc}$ are the number of successful trials, the total number of trials, and the success probability for the binomial distribution respectively.
Then, the differential rate over $n_e$
is given by
\begin{align}
{\td \mathcal{R}_{{\rm ion}}^{n \ell}\over \td n_e}
=\int {\td \ln E_e} 
{\td \mathcal{R}_{{\rm ion}}^{n \ell}\over \td \ln E_e} P_{\rm ion}^{n\ell}(n_e|E_e).
\end{align}
In our analysis, we adopt $f_R = 0$ and $f_e = 0.83$ \cite{Essig:2012yx}.

For each ionized electron, it is assumed that the number of S2 photoelectrons (PEs), $N_{\rm PE}$, induced by it follows a Gaussian distribution with a mean value $g_2$ and a width value $\sigma_{\rm S2}$. 
For XENON10 (XENON1T), $g_2=27\,(33)$  and $\sigma_{\rm S2} = 7$ \cite{XENON10:2011prx,XENON100:2013wdu,XENON:2019gfn}. The parameters $g_2$ and $\sigma_{\rm S2}$ for the PandaX-4T S2-only search were not given explicitly in \cite{PandaX:2022xqx}. 
We inferred from its Fig.\,(3) that $g_2=17.7$ and assume $\sigma_{\rm S2}=\sqrt{g_2}$.
For multiple ionized electrons, the probability that $n_e$ electrons produce $N_{\rm PE}$ S2 PEs is given by:
\begin{align}
P( N_{\rm PE} | n_e) = {1\over \sqrt{2\pi} \sqrt{n_e} \sigma_{\rm S2} } \exp\left[- {1\over 2} 
\left({N_{\rm PE}- g_2 n_e\over \sqrt{n_e}\sigma_{\rm S2}}\right)^2\right].
\end{align}
Finally, we obtain the differential rate of the number of signal events over $N_{\rm PE}$:
\begin{align}
{\td N_s \over \td N_{\rm PE}}=  \epsilon \, \omega \,
{1 \over m_T} \sum_{n,\ell} \sum_{n_e=1}^{\infty} 
{\td \mathcal{R}_{{\rm ion}}^{n \ell}\over \td n_e} P( N_{\rm PE} | n_e),
\label{eq:dNdPE}
\end{align}
where $m_T$ is the mass of the xenon atom, $\omega$ is the exposure of the experiment, and $\epsilon$ accounts for efficiency factors, including trigger efficiency and selection efficiencies. For the sum over $n,\ell$, we consider the five outermost occupied orbitals of the xenon atom (4s, 4p, 4d, 5s, 5p). For the sum over $n_e$, we set a truncation with $n_e \leq 15$ in the numerical calculation.

Here we provide the criteria to set 90\,\% C.L. constraints from the S2-only data in XENON10, XENON1T, and PandaX-4T experiments.
\\
\indent%
\textbf{XENON10:}
The criterion of 90\,\% C.L. constraints for XENON10 is given in \cite{Essig:2017kqs}.
For the seven event bins $N_{\rm PE}^i \in [27i-13,27i+14] $ with $i=1,2,\dots,7$, the corresponding upper bounds on the event rates (after unfolding the efficiencies) are $r_1< 15.18, r_2< 3.37, r_3< 0.95, r_4< 0.35, r_5< 0.35, r_6< 0.15, r_7< 0.35$ counts $\rm kg^{-1}day^{-1}$ respectively, where 
$r_i = \int_{27i-13}^{27i+14} \td N_{\rm PE} 
{1\over \epsilon\,\omega} {\td N_s \over \td N_{\rm PE}}$. 
\\
\indent%
\textbf{XENON1T:}
For the event bin $N_{\rm PE}$ $\in [165,271]$, the 90\,\% C.L. constraint requires that 
the number of signals is less than 24.7 \cite{XENON:2019gfn}.
We use the exposure of 80755 kg days, and consider all kinds of efficiencies given in Fig.\,(2) of \cite{XENON:2019gfn} and the combined efficiency of 93\,\%.
\\
\indent%
\textbf{PandaX-4T:}
With the exposure of 0.55 ton$\cdot$yr, 103 events (including 95.8 background events estimated) were observed in the signal region of  $ N_{\rm PE}\in [60,200]$ \cite{PandaX:2022xqx}.
According to the Poisson distribution, the 90\,\% C.L. requires the number of signal events to be less than 21.5.
We employ the total efficiency (red solid line) given in Fig.\,(1) of \cite{PandaX:2022xqx}.

\subsection{Results}

In our analysis, each constraint is evaluated by considering the contribution of only a single operator.
For the relativistic EFT operators, we parameterize their dimensional WCs in terms of some effective scale $\Lambda$ to set the bound, i.e., $C_i \equiv 1/\Lambda^n$ with $n$ being determined by the dimension of the operator $\calO_i$ under consideration.
In general, the PandaX-4T data set the most stringent constraints for the DM mass region of $m_x \gtrsim 20 \, \rm MeV$. 
For the low mass region of $m_x \lesssim 20 \, \rm MeV$, the constraints from XENON10 experiment remain leading.
For most operators, except for certain operators in the vector DM case, the most stringent constraints occur at $m_x \simeq 0.1 \, \rm GeV$. 
For $m_x \lesssim 0.1 \, \rm GeV$, the constraints are limited by the detection threshold of the experiments; for $m_x \gtrsim 0.1 \, \rm GeV$, the constraints are weakened due to the lower DM number density for heavier DM.

\subsubsection{Non-relativistic EFT interactions}

\begin{figure}
\centering
\includegraphics[width=0.99 \textwidth]{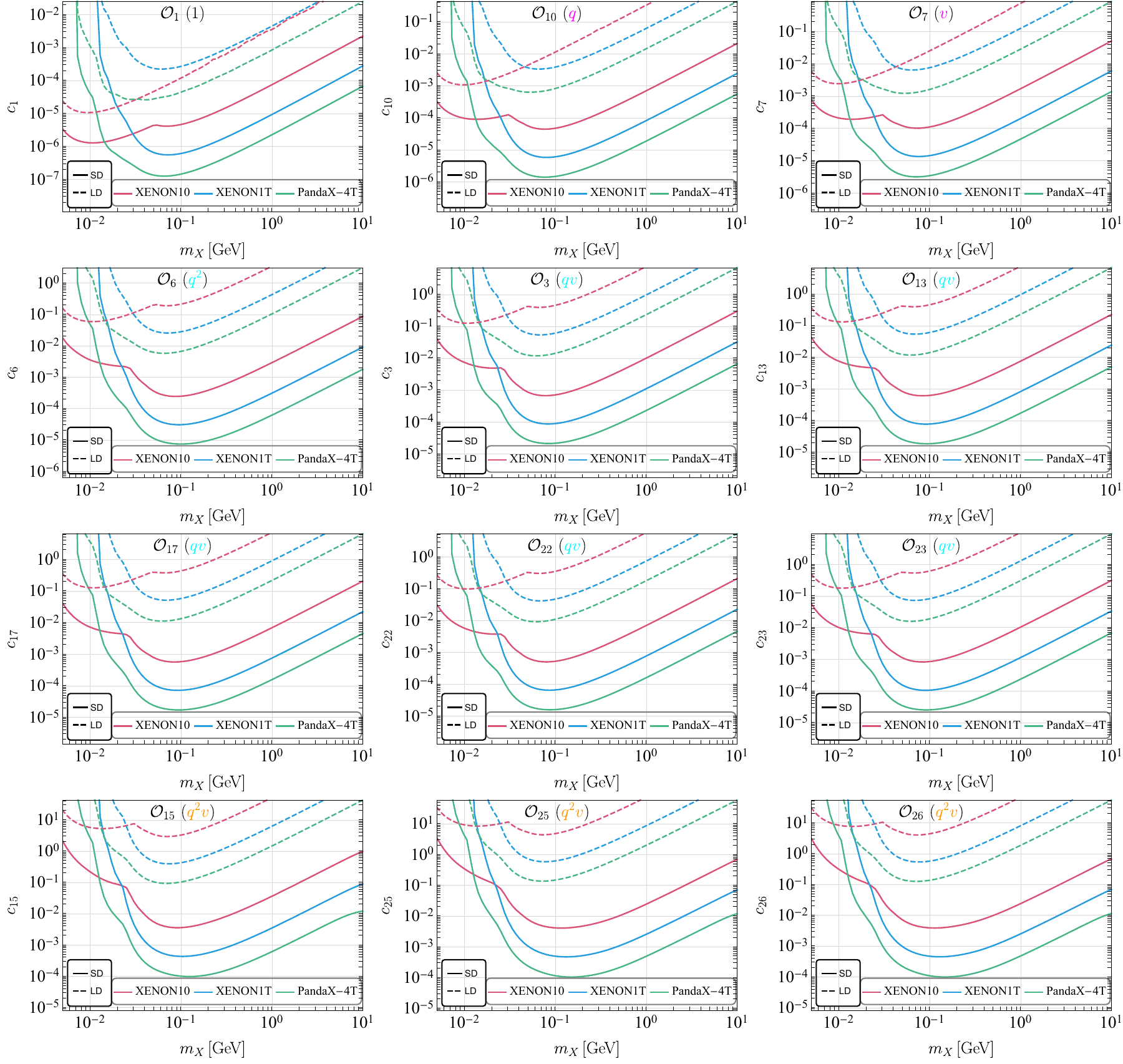}
\caption{90\,\% C.L. constraint on each benchmark WC in \cref{eq:bmWC} for vector DM in the SD (solid curves) and LD (dashed curves) cases from XENON10, XENON1T, and PandaX-4T, where the symbols in the parentheses indicate the corresponding power counting.}
\label{fig:vec-con}
\end{figure}

In \cref{fig:vec-con}, we present the 90\,\% C.L. constraints on the WCs for the benchmark vector operators listed in \cref{tab:rescal} and \cref{eq:bmWC}.
Both the SD (solid curves) and LD (dashed curves) cases are considered.
For the LD case, one has to replace $c_i$ by $c_i\, q_{\rm ref}^2/\bm{q}^2$ in \cref{tab:dmRF}, with $q_{\rm ref} = \alpha_{\rm em} m_e$ being a typical reference momentum.
The constraints for the scalar, fermion, and remaining vector operators can be 
simply obtained by multiplying the corresponding rescaling factors shown in the third-to-fifth columns in \cref{tab:rescal}.
In \cref{fig:vec-con}, we also display the power counting for each NR operator and organize them accordingly.  
The constraints on operators with an identical power counting are nearly degenerate. For instance, the constraints on $\calO_3$, $\calO_{13}$, $\calO_{17}$, $\calO_{22}$, and $\calO_{23}$, which all share the same $qv$ power, are similar. 
Their WCs appear in $a_{1,2}$ and are accompanied by a factor of $x_e$, together with a similar dependence on the atomic response functions $\widetilde W_{1,2}$ (c.f.\,\cref{tab:rescal}), leading to the similar constraints.\footnote{For $\calO_3$ and $\calO_{15}$, the combination of atomic response functions in \cref{tab:rescal} takes the approximate form, $\widetilde W_1 - \widetilde W_2 \approx W_3 - W_4/x_e$, whose size is comparable with $x_e W_1$.}
When comparing constraints on $\calO_{10}$ and $\calO_7$, it is evident that the power of $v$ results in slightly stronger suppression than the power of $q$. 
Approximately, the constraint on $c_{10}$ is roughly twice stronger than that of $c_7$ in each experiment. 
This can be quantitatively understood as follows. 
Since the correction of $\Delta E_{1\to 2}/m_e$ in \cref{eq:v02andye} only reaches up to $40\,\%$ relative to the leading $x_e$ term, neglecting its contribution is still a good approximation to simplify $\widetilde W_{1,2}$, and we find $\widetilde W_1 \approx x_e W_1/4 + W_2 + W_3$ and $\widetilde W_2 \approx x_e W_1/4 + W_2 +  W_4/x_e$. 
These approximate relations and the cancellation behavior between $W_2$ and $W_3$ \cite{Liang:2024lkk} further lead to $\widetilde W_1 \approx x_eW_1/4$.
From \cref{tab:rescal}, the matrix element squared due to the operator $\calO_{10}$ is $c_{10}^2 \times x_e\widetilde W_0/4 \simeq c_{10}^2 (x_e W_1/4)$ while for $\calO_7$ it becomes $c_7^2 \times \widetilde W_1/4\approx (c_7/2)^2 (x_e W_1/4)$, leading to a factor of 2 difference in the constraints. A similar situation happens for $c_6$ and $c_{13}$ as can be seen from \cref{tab:rescal}, but with an additional factor of $2x_e/3$ for each operator.
On the other hand, by approximating $q$ and $v$ as being of the same order, we observe that the constraints on the WCs are weakened by about an order of magnitude with each additional power of $q$ or $v$ for the operators, a number that can be estimated by using $m_e/q \approx 10\,(50\,{\rm keV}/q)$ for a typical momentum transfer $q\sim 50\,\rm keV$.
Generally, for the same operator and experiment, the difference of the constraints between the SD and LD cases runs from about one to two orders of magnitude as the DM mass increases from 5\,MeV to 100\,MeV. This can be understood from the definition of the LD case in which there is an additional suppression factor for the WC, $(q/(\alpha_{\rm em} m_e))^2\sim 180(q/50\,\rm keV)^2$.

\subsubsection{Relativistic EFT interactions}

For a relativistic operator that induces a single NR operator, one can directly obtain the constraint on it according to the matching relations offered in \cref{tab:NRmatch:sf}.
For a relativistic operator that induces several NR operators, we need to combine the matching relations offered in \cref{tab:NRmatch:sf,tab:NRmatch:v}, and the general DM response functions given in \cref{tab:dmRF} to calculate the constraints so that interference effects among NR operators are automatically incorporated.

\begin{figure}
\centering
\includegraphics[width=0.99 \textwidth]{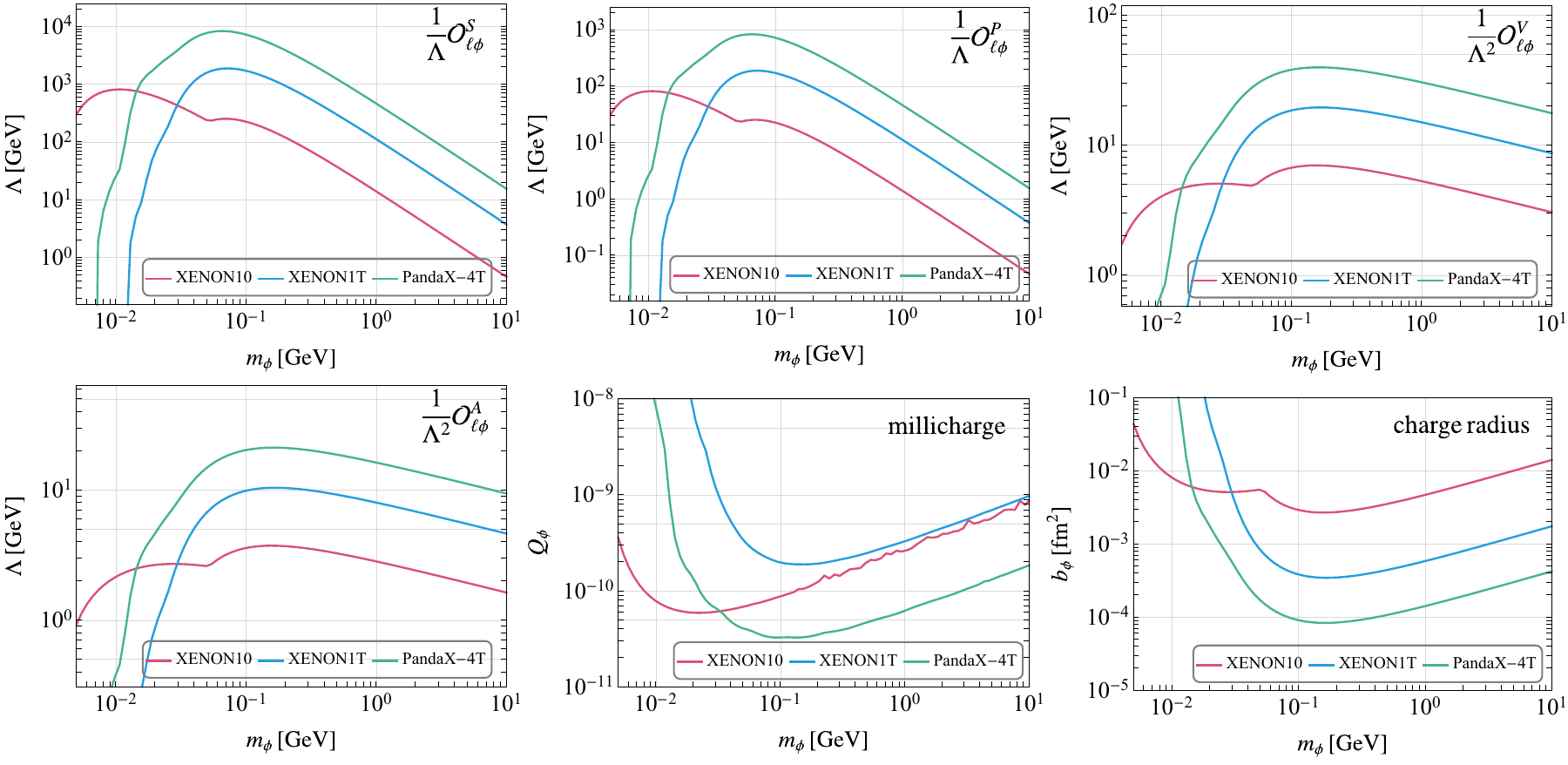}
\caption{90\,\% C.L. constraints on the relativistic EFT operators for the scalar DM.}
\label{fig:vec-sca}
\end{figure}

\pmb{Scalar DM case:}
The constraints on relativistic operators for scalar DM are shown in \cref{fig:vec-sca}. Since all scalar relativistic operators have only one NR operator contribution at LO expansion, as shown in \cref{tab:NRmatch:sf}, the constraints on the scalar operators can be directly obtained by rescaling the constraints on NR operators.
Only the millicharge interaction term $\calL_{\phi}^{Q}$ can induce a LD NR contribution.
As expected, the constraints on the effective scale for dim-5 operators ($\calO_{\ell \phi}^S$ and $\calO_{\ell \phi}^P$) are much stronger than those for dim-6 operators ($\calO_{\ell \phi}^V$ and $\calO_{\ell \phi}^A$), by two to three orders of magnitude.
The constraints on $\calO_{\ell \phi}^S$ are approximately one order of magnitude stronger than those on $\calO_{\ell \phi}^P$,  
with the most stringent constraint on the former reaching $\Lambda \gtrsim 10\, \rm TeV$. This distinction arises because $\calO_{10}$, which is induced by $\calO_{\ell \phi}^P$, has a dependence on $q$, resulting in an additional suppression.
Without the factor of $m_\phi$ in the NR reduction,
the constraints on $\calO_{\ell \phi}^{S,P}$ drop quickly than those of $\calO_{\ell \phi}^{V,A}$ at large $m_\phi$.
The most stringent constraint on the charge (charge radius) for the scalar DM is found at $ m_\phi \simeq 0.1\, \rm GeV $, yielding $ Q_\phi \lesssim 3\times 10^{-11} $ ( $ b_\phi \lesssim 10^{-4}\, \rm fm^2 $), from the latest PandaX-4T data.

\begin{figure}
\centering
\includegraphics[width=0.99 \textwidth]{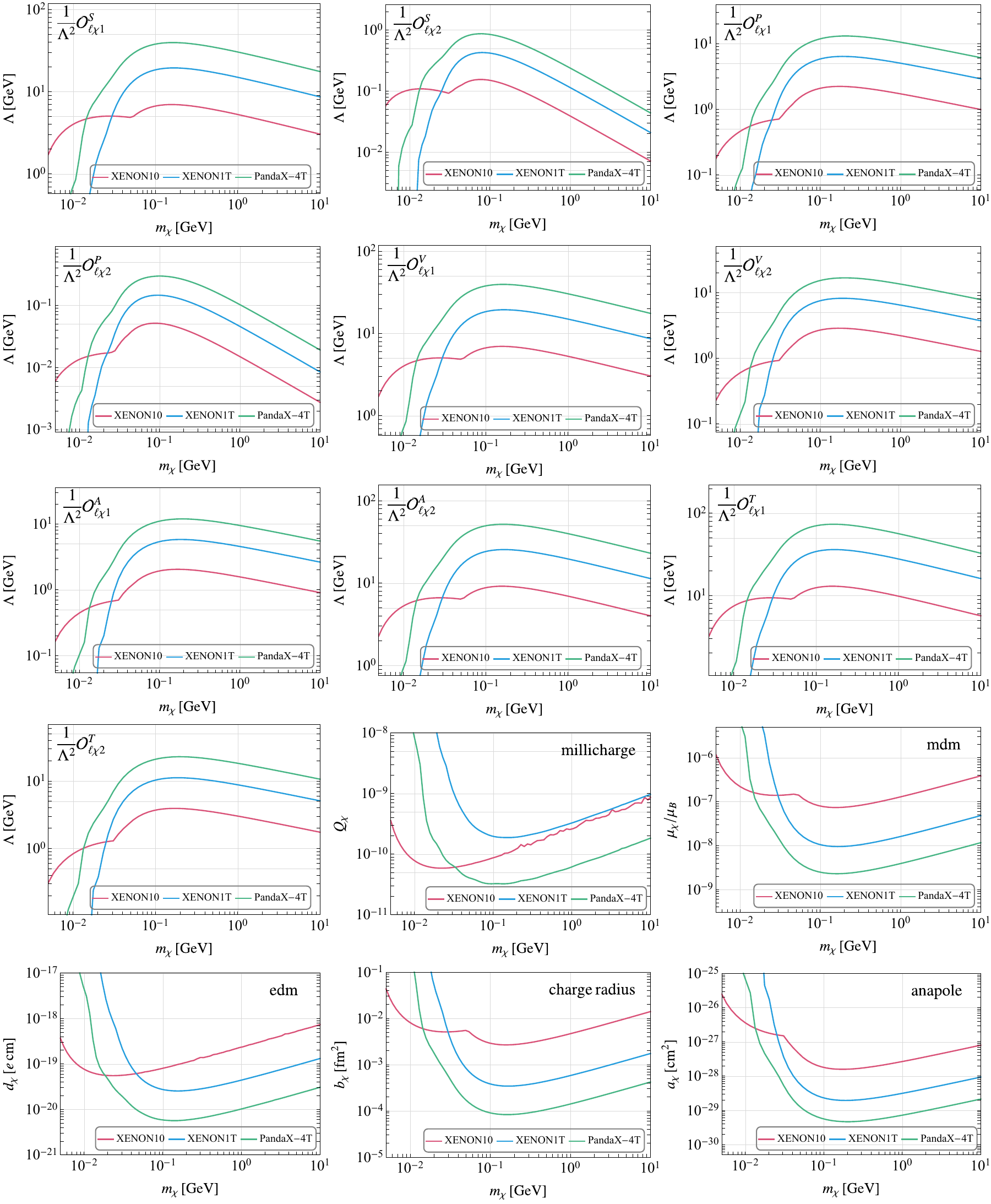}
\caption{90\,\% C.L. constraints on the relativistic EFT operators for the fermion DM. }
\label{fig:vec-fer-All}
\end{figure}

\pmb{Fermion DM case:}
The constraints on relativistic operators for fermion DM are shown in \cref{fig:vec-fer-All}.
The operators, $\calO_{\ell \chi 2}^V$, $\calO_{\ell \chi 1}^A$, $\calO_{\ell \chi 2}^T$,
and interactions
$\calL_\chi^{\rm mdm}$, $\calL_\chi^{\rm anap.}$, all involve  interference effects among NR operators.
The three EM interactions, $\calL_\chi^{Q}$, $\calL_\chi^{\rm mdm}$, and $\calL_\chi^{\rm edm}$, can induce LD NR interactions.
Except for $\calO_{\ell \chi 2}^S$ and $\calO_{\ell \chi 2}^P$, the other four-fermion operators include a factor of $m_\chi$ in their NR reduction results. This  results in stronger constraints and a more gradual decrease in the constraints for $m_\chi \gtrsim 0.1\,\rm GeV$,
with the constraints being $\Lambda \gtrsim 10-80 \,\rm GeV$.
Without the enhancement from the DM mass factor, 
$\calO_{\ell \chi 2}^S$ and $\calO_{\ell \chi 2}^P$ are poorly constrained, with $\Lambda \gtrsim 0.1-1 \,\rm GeV$.
For the electromagnetic properties of the fermion DM,
the most stringent constraints reach $Q_\chi\lesssim 3\times 10^{-11}$,
$\mu_\chi /\mu_B \lesssim 2\times 10^{-9}$, $d_\chi \lesssim 7 \times 10^{-21} e \,\rm cm$, $b_\chi \lesssim 10^{-4} \rm \,fm^2$, and $a_\chi \lesssim 6 \times 10^{-30} \,\rm cm^2$, respectively.  Here $\mu_B$ is the Bohr magneton.

\begin{figure}
\centering
\includegraphics[width=0.99 \textwidth]{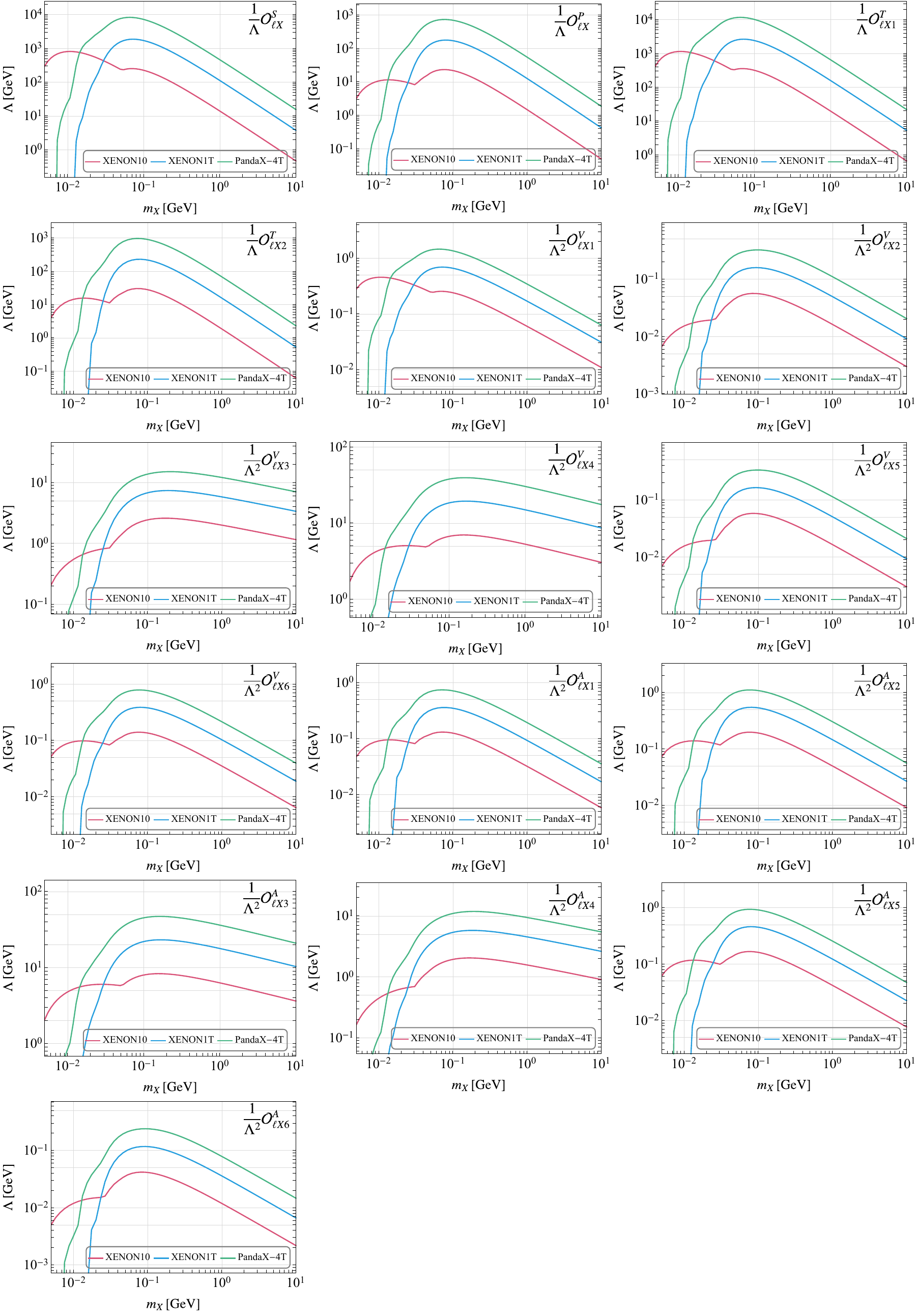}
\caption{90\,\% C.L. constraints on case-A vector relativistic EFT operators.}
 \label{fig:vec-rel-A}
\end{figure}
\begin{figure}
    \centering
 \includegraphics[width=0.99 \textwidth]{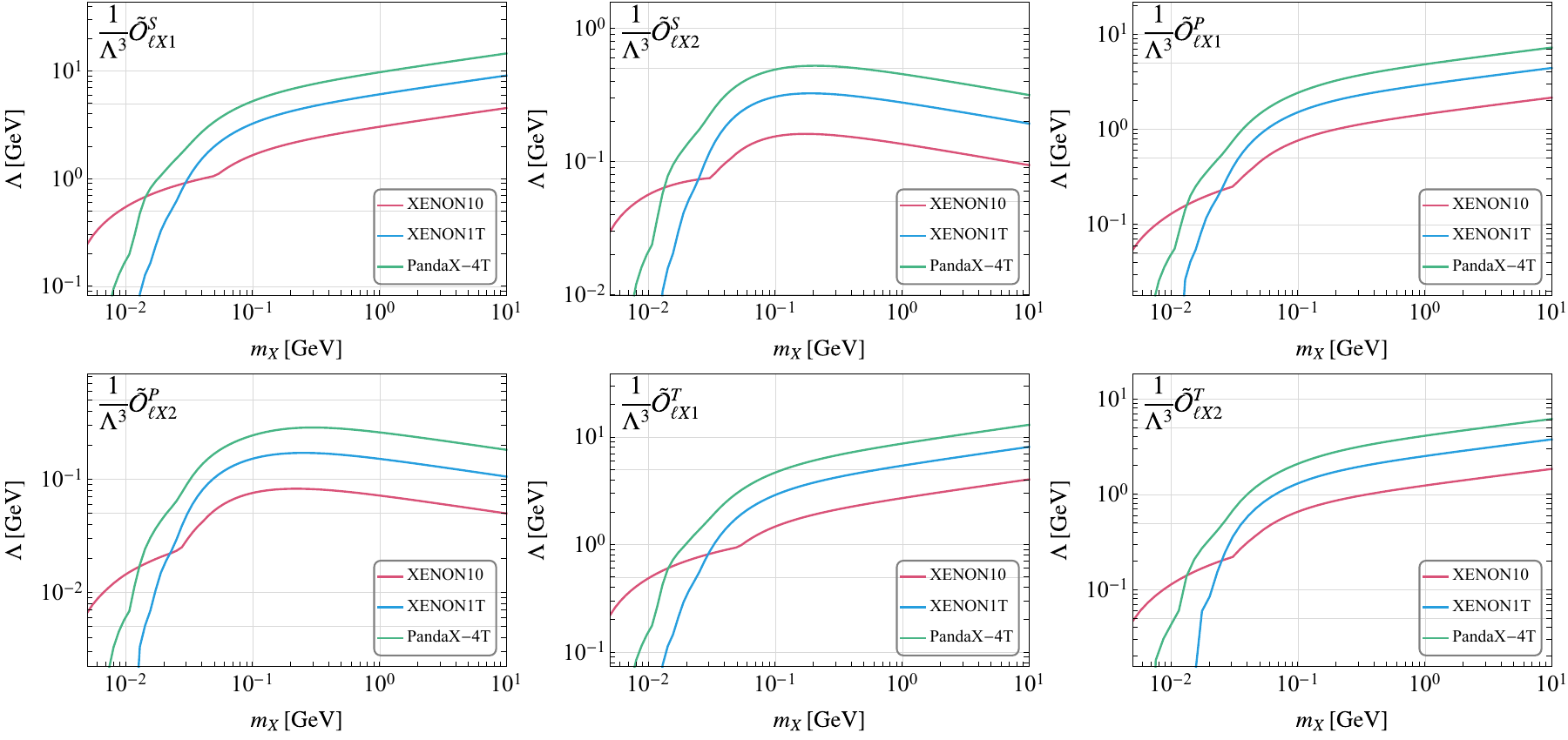}
    \caption{90\,\% C.L. constraints on case-B vector relativistic EFT operators}
    \label{fig:vec-rel-B}
\end{figure}
\begin{figure}
    \centering
 \includegraphics[width=0.99 \textwidth]{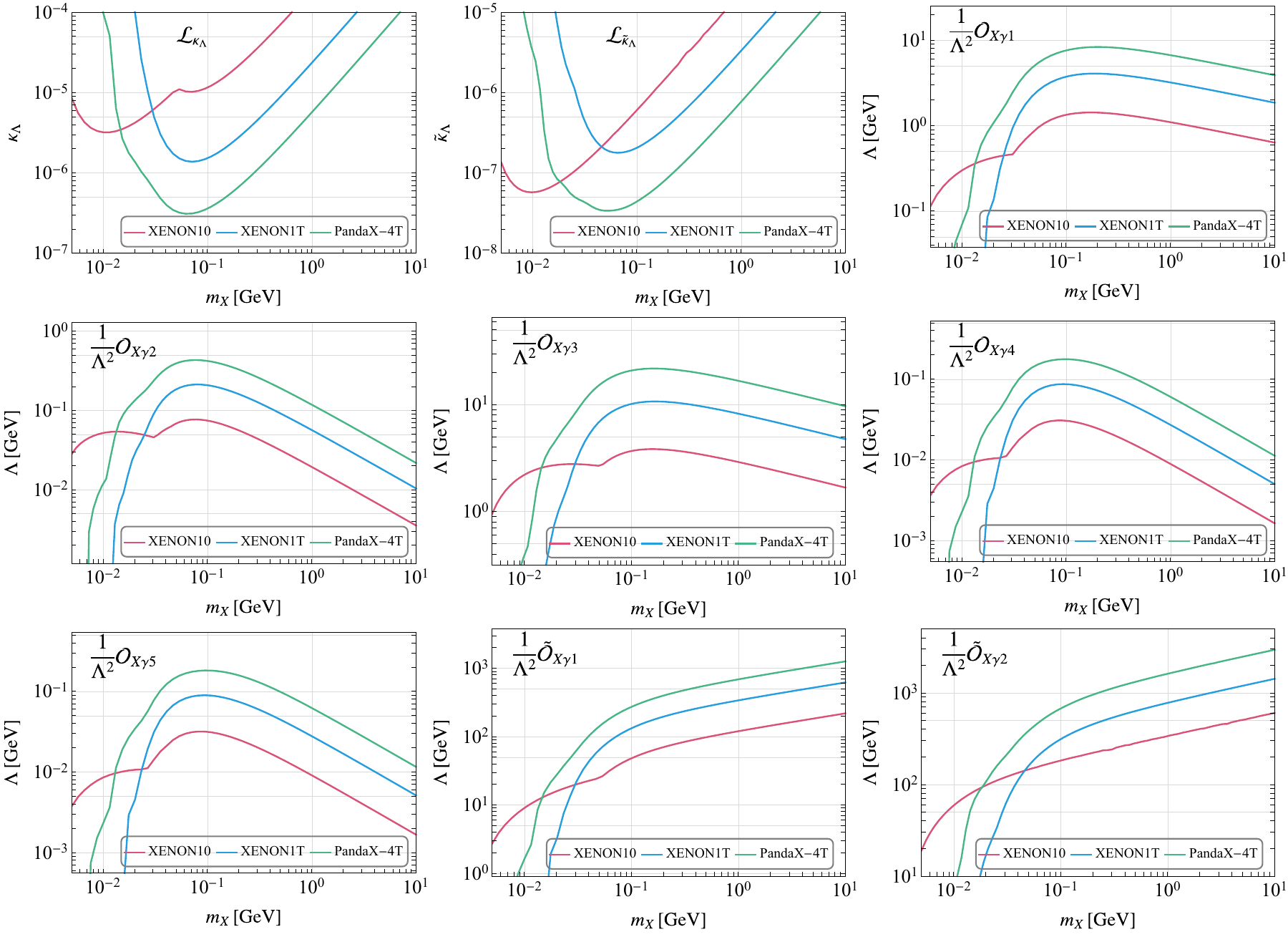}
    \caption{90\,\% C.L. constraints on relativistic EM operators for vector DM.
    }
    \label{fig:vec-rel-EM}
\end{figure}

\pmb{Vector DM SD - case A:}
The constraints on the SD case A operators are shown in  
\cref{fig:vec-rel-A}.
The four dim-5 operators fall into two groups with comparable constraints: $\calO_{\ell X}^S$ and $\calO_{\ell X 1}^T$ have constraints reaching $\Lambda \gtrsim 10 \,\rm TeV$, whereas $\calO_{\ell X}^P$ and $\calO_{\ell X 2}^T$ are weaker by about one order of magnitude.
As for the dim-6 operators, they are poorly constrained except for $\calO_{\ell X 3}^V$, $\calO_{\ell X 4}^V$, $\calO_{\ell X 3}^A$, and $\calO_{\ell X 4}^A$. However, the constraints on these operators still remain weak, with $\Lambda \gtrsim 10 \, \rm GeV$.

\pmb{Vector DM SD - case B:}
The constraints on the SD case B operators are shown in \cref{fig:vec-rel-B}.
These operators are all at dim 7, hence the constraints on them are relatively weak, with the most stringent constraints only reaching $\Lambda \gtrsim 10\, \rm GeV$.
Special behavior is observed for $\tilde{\calO}_{\ell X 1}^S$, $\tilde{\calO}_{\ell X 1}^P$, $\tilde{\calO}_{\ell X 1}^T$, and $\tilde{\calO}_{\ell X 2}^T$, with the constraints monotonically increasing with the mass.
This behavior arises from the $ m_X^2 $ factor in their NR matching, which counteracts the suppression resulting from the DM number density in the high DM mass region.

\pmb{Vector DM EM:}
The constraints on the EM operators for vector DM are shown in  
\cref{fig:vec-rel-EM}.
The constraints on $\tilde{\calO}_{X\gamma1}$ and $\tilde{\calO}_{X\gamma1}$ monotonically increase with the mass, reaching $\Lambda \gtrsim 1 \, \rm TeV$ at $m_X \simeq 10 \,\rm GeV$. These are the strongest constraints in this large DM mass region, compared to all other operators, including scalar, fermion, and vector DM.
For two dim-4 interaction terms, the constraint on $\kappa_{\Lambda}$ ($\tilde{\kappa}_{\Lambda}$) can reach about $2\times 10^{-7}$ ($3\times 10^{-8}$).

\section{Summary}
\label{sec:conclusion}

In this paper, we have systematically investigated the general dark matter-electron interactions within the framework of effective field theories (EFT). We considered both the non-relativistic and relativistic EFT descriptions of the interactions with the spin of dark matter (DM) up to one, i.e., the scalar DM ($\phi$), the fermion DM ($\chi$), and the vector DM $(X)$ scenarios. We provided the leading-order NR EFT operators describing the DM-electron interactions, especially the construction of NR operators for the vector DM case. 
To study potential origin of the NR operators, we then explored all possible leading-order relativistic DM-electron and DM-photon interactions in the LEFT-like framework, and performed the NR reduction to match them onto the NR EFT.
These results can be readily extended to the SMEFT-like framework or UV models. To make connections with experimental searches, we then
rederived the DM-bound-electron scattering rate within the NR EFT framework, including all leading order effective operators for the three (scalar, fermion, and vector) DM scenarios. We found that the DM-atom scattering matrix element squared, which correlates the DM and atomic properties, can be compactly decomposed into three terms. Each term is a product of a DM response function $(a_{0,1,2})$ and its corresponding generalized atomic response function ($\widetilde W_{0,1,2}$).
Our formalism for DM-atom scattering makes it manifest which type of effective interactions is associated with which type of atomic response functions (see \cref{eq:W012tilde} and \cref{tab:dmRF}). More concretely, $a_0$ (associated with $\widetilde{W}_0$) is exclusively relevant to velocity-independent operators while the velocity-dependent ones are captured by $a_{1,2}$ (associated with $\widetilde{W}_{1,2}$), and all of $a_{0,1,2}$ have a definite power counting in $x_e$. This organization of terms is very useful to quickly identify the primary contributions to the DM-atom scattering within the framework of relativistic EFT and UV complete models.
Finally, we utilized electron recoil data from DM direct detection experiments with a xenon target, including XENON10, XENON1T, and PandaX-4T, to constrain all non-relativistic and relativistic EFT interactions across all three DM scenarios. Our results indicate that the latest PandaX-4T S2-only data generally impose more stringent constraints on the EFT interactions of dark matter with a mass greater than approximately 20 MeV, surpassing those from earlier XENON10 and XENON1T experiments.

For the NR EFT operators, we presented explicitly the constraints on the dimensionless WCs of the 12 benchmark vector DM operators, while the constraints on all other NR operators, including those for fermion and scalar DM, can be obtained by rescaling based on the relations in \cref{tab:dmRF} and \cref{tab:rescal}. 
These constraints generally exhibit an upward-opening parabolic shape, with the lowest point at $m_x \simeq 0.1 ~\rm GeV$ ranging from approximately $10^{-4}$ to $10^{-7}$. 
The constraints generally follow the power counting rules in $q$ and $v$, with the strongest being on $\calO_1$ (counting order one) and the weakest on operators such as $\calO_{15}$, $\calO_{25}$, and $\calO_{26}$ (counting order $q^2v$).
Additionally, we observed that the velocity $v$ induces a slightly stronger suppression than the momentum transfer $q$.

Concerning relativistic EFT operators we generally displayed the constraints on the effective scale for each operator while for the interactions associated with the electromagnetic properties of DM, we showed the constraints on the millicharge, the electric and magnetic dipole moments, the anapole, and the charge radius, respectively.
The most stringent constraints on the effective scale come from the three dim-5 operators ($\calO_{l\phi}^S$, $\calO_{lX}^S$, and $\calO_{lX1}^T$), which can induce the NR operators with the leading power counting ($\calO_1$ and $\calO_4$). They can reach about 10 TeV for DM with a mass of about 0.1 GeV. 
Specifically, the constraints on some vector DM operators ($\tilde{\calO}_{\ell X 1}^S$, $\tilde{\calO}_{\ell X1}^P$, $\tilde{\calO}_{\ell X 1}^T$, $\tilde{\calO}_{\ell X 2}^T$, $\tilde{\calO}_{X\gamma 1}$, and $\tilde{\calO}_{X\gamma 2}$ in the case B) become stronger as the DM mass increases, despite of the suppression from the decrease in the DM number density.

\acknowledgments
This work was supported in part by the Grants 
No.\,NSFC-12035008, 
No.\,NSFC-12247151, 
No.\,NSFC-12305110,
and No.\,NSFC-12347121,
and by the Guangdong Major Project of Basic
and Applied Basic Research No.\,2020B0301030008.

\appendix

\section{Details for constructing NR operators involving spin tensor operator }
\label{app:NRop}

In this appendix, we provide the details for constructing NR operators involving the rank-two tensor operator $\pmb{\tilde\calS}_x$ in the case of vector DM-electron scattering. To maintain rotational invariance, another rank-two tensor $T^{ij}$ should be formed to contract with ${\Tilde{\calS}}_x^{ij}$ so that the obtained NR operators are scalar objects under the space rotation group. Since ${\Tilde{\calS}}_x^{ij}$ is constructed to be a symmetric tensor under the exchange of $i\leftrightarrow j$, the anti-symmetric component of $T^{ij}$ has no contribution. From the available building blocks, $\{\bm{q}, \pve\}\otimes\{\mathbb{1}_e,\pSe\}$, one way to build $T^{ij}$ is by choosing two vector objects from this set and arranging them in the form $A^i B^j$. In this way, there are four independent LO operators, 
\begin{subequations}
\begin{align}
\calO_{17} &= {i \bm{q} \over m_e} \cdot \pSxt \cdot \pve,
\\
\calO_{18} &=  {i \bm{q}\over m_e} \cdot \pSxt \cdot  \pSe,
\\
\calO_{19} &=  {\bm{q} \over m_e} \cdot \pSxt \cdot {\bm{q}\over m_e} \mathds{1}_e,
\\
\calO_{21} &= \pve\cdot\pSxt\cdot\pSe.
\end{align}
\end{subequations}
The other way to construct $T^{ij}$ is through a vector object and a cross product of two vector objects in the form, $A^i(\bm{B}\times\bm{C})^j$. the possible structures are,
\begin{align}
\label{eq:struc}
q^i (\bm{q}\times \pve)^j ,\quad q ^i(\bm{q}\times \pSe)^j ,\quad S_e^i (\bm{q}\times \pve)^j,\quad v_{\rm el}^{\perp i}(\bm{q}\times \pSe)^j,\quad q^i(\pve\times \pSe)^j.
\end{align}
Due to the Schouten identity \cite{Liao:2020jmn}, 
\begin{align}
\delta^{ij}\epsilon^{klm}-\delta^{ik}\epsilon^{lmj}+\delta^{il}\epsilon^{mjk}-\delta^{im}\epsilon^{jkl}=0,
\end{align}
the above structures are not independent, but satisfy the following relation, 
\begin{align}
q^i(\pve\times\pSe)^j-\delta^{ij}\bm{q}\cdot (\pve\times\pSe)-\pve^i (\bm{q}\times\pSe)^j +  S_e^i (\bm{q}\times \pve)^j =0.
\label{eq:SI}
\end{align}
Contracting \cref{eq:SI} with ${\Tilde{\calS}}^{ij}_x$ and in view of ${\Tilde{\calS}}^{ij}_x$ being traceless, we find that only two out of the last three structures in \cref{eq:struc} are independent upon contracting with ${\Tilde{\calS}}_x^{ij}$. We therefore choose the following four operators to form the basis:
\begin{subequations}
\begin{align}
\calO_{20} &=  - {\bm{q}\over m_e}  \cdot \pSxt \cdot \left({\bm{q}\over m_e} \times \pSe\right),
\\
\calO_{22} &= \left({i \bm{q} \over m_e}\times\pve\right)\cdot\pSxt\cdot\pSe + \pve\cdot\pSxt\cdot\left({i \bm{q} \over m_e}\times\pSe\right),
\\
\calO_{23} &= - {i \bm{q} \over m_e}\cdot\pSxt\cdot(\pve \times \bm{S}_e),
\\
\calO_{24} &= {\bm{q} \over m_e} \cdot \pSxt \cdot  \Big({ \bm{q} \over m_e} \times\pve\Big).
\end{align}
\end{subequations}
The last possibility to compose $T^{ij}$ is by using four vector objects with two being contracted as a scalar, i.e., $T^{ij}=(\bm{A}\cdot\bm{B})\,C^i D^j$. In this case, there are at most two possibilities at second order in $\bm{q}$ and linear order in $\pve$:
\begin{subequations}
\begin{align}
\calO_{25}  &= \Big({\bm{q}\over m_e} \cdot \pSxt\cdot\pve \Big) 
\Big({\bm{q}\over m_e} \cdot\bm{S}_e \Big),
\\
\calO_{26} &= \Big({\bm{q}\over m_e} \cdot \pSxt \cdot {\bm{q}\over m_e} \Big) (\pve\cdot\bm{S}_e).
\end{align}
\end{subequations}
The operator formed with a scalar $\bm{q}\cdot\pve$ is not independent and thus not included since $\bm{q}\cdot\pve$ is a trivial kinematic variable while $\bm{q}\cdot \pSxt \cdot \bm{S}_e $ is already counted. In total, there are 10 more operators associated with $\pSxt $.
Notice that the two operators $\calO_{21}$ and $\calO_{22}$ have no contribution to the nuclear recoil signal for the DM-nucleon scattering in the one-nucleon current approximation \cite{Gondolo:2020wge}.

\section{Calculation of DM response functions}
\label{app:dmRF}

To calculate the matrix element squared, we rewrite the transition amplitude  \cref{eq:M1to2new} as 
\begin{align}
  \calM_{1\to 2} = f_{\tt S} \calM_{\tt S}  + {\bm f}_{\tt V}\cdot {\bm \calM}_{\tt V}
  \equiv \eta_{\mu\nu}\tilde F^\mu \tilde\calM^\nu,  
  \quad 
  \tilde F^\mu = (f_{\tt S}, {\bm f}_{\tt V}), \quad
  \tilde\calM^\mu = ( \calM_{\tt S} , {\bm \calM}_{\tt V}), 
\end{align}
where we have introduced the four-vectors ($\tilde F^\mu,\tilde\calM^\mu$) and the Euclidean metric $\eta_{\mu\nu}\equiv {\rm diag}(1,1,1,1)$ with $\mu,\nu=0,1,2,3$. The introduction of this formalism is quite useful to handle the non-relativistic quantities in a 
pseudo-relativistic fashion using the Euclidean metric. The spin-averaged transition amplitude squared takes the form
\begin{align}
 \overline{|\calM_{1\to2}|^2} = 
 \tilde F_\mu^* \tilde F_\nu\,
 R^{\mu\nu}, \quad 
 R^{\mu\nu} \equiv {1\over s_x }{1\over s_e} \sum_{\rm pol.} \tilde\calM^{\mu*} \tilde\calM^\nu, 
 \label{eq:Rmunu}
\end{align}
where $s_x=1,2,3$ are respectively the number of spin states of the scalar, fermion, and vector DM, while $s_e=2$ is the spin states of the electron.

Since the NR interactions generally take a current-current factorized form in terms of the following DM and electron NR current operators,
\begin{align}
j_x = 
\begin{pmatrix}
\mathbb{1}_x \\   S_x^i \\   \tilde{\calS}_x^{ij}
\end{pmatrix}, 
\quad 
j_e = 
\begin{pmatrix}
\mathbb{1}_e \\ S_e^k
\end{pmatrix}, 
\end{align}
where the indices $i,j$ (and $n,m$ in the following) are used for the DM spin component while $k$ ($l$) the electron's spin component. 
In terms of $j_x$ and $j_e$, the DM-free-electron scattering amplitude from the initial state 
$|\bm{p}, \lambda; \bm{k}, s\rangle$ to the final state $|\bm{p}', \lambda'; \bm{k}', s'\rangle$ can be written as a product of matrices capturing different physics, 
\begin{align}
\tilde \calM^\nu  =  
\langle \lambda'| j_x^\dagger | \lambda \rangle\tilde C_{x e}^\nu  \langle s'| j_e | s \rangle  
= \langle \lambda'|
\begin{pmatrix}
\mathbb{1}_x & 
 S_x^i  &
\tilde{\calS}_x^{ij}
\end{pmatrix}
|\lambda \rangle 
\begin{pmatrix}
\tilde C_{xe,11}^\nu & \tilde C_{xe,12}^{k,\nu}  
\\
\tilde C_{xe,21}^{i,\nu} & \tilde C_{xe,22}^{i,k,\nu} 
\\
\tilde C_{xe,31}^{i j,\nu} &\tilde C_{xe,32}^{ij,k,\nu} 
\end{pmatrix}
\langle s'|
\begin{pmatrix}
\mathbb{1}_e\\ 
S_e^k
\end{pmatrix}
|s  \rangle,
\end{align}
and the complex conjugate amplitude 
\begin{align}
\tilde \calM^{\mu*} 
 = 
\langle s| j_e^\dagger|s'\rangle\tilde C_{xe}^{\mu\dagger} \langle \lambda |j_x|\lambda' \rangle  
= 
\langle s |
\begin{pmatrix}
\mathbb{1}_e & 
 S_e^l 
\end{pmatrix}
| s' \rangle 
\begin{pmatrix}
\tilde C_{xe,11}^{\mu*} &\tilde  C_{xe,21}^{m,\mu *} &\tilde C_{xe,31}^{mn,\mu *} 
\\
\tilde C_{xe,12}^{l,\mu *} &\tilde C_{xe,22}^{m,l,\mu *} & \tilde C_{xe,32}^{mn,l,\mu *} 
\end{pmatrix}
\langle \lambda|
\begin{pmatrix}
\mathbb{1}_e \\ S_X^m \\ \tilde{\calS}_X^{mn}
\end{pmatrix}
|\lambda'  \rangle. 
\nonumber
\end{align}
In the above the linking matrix $\tilde C_{xe}^\nu $ can be directly read off from the NR operators together with their corresponding WCs. 
For instance, the $\calO_1$ operator leads to $\tilde C_{xe,11}^{\nu} = c_1 \eta_{\nu 0}$
while for $\calO_3$,  $\tilde C_{xe,12}^{k,\nu} = c_3 \epsilon^{k ij} {i\,q^i \over m_e}\eta_{\nu j}$, etc. Thus, the $R^{\mu\nu}$ defined in \cref{eq:Rmunu} is simplified to become a trace of matrices, 
\begin{align}
R^{\mu\nu}
= {1\over s_x s_e }  \sum_{\rm pol.} 
\langle s|j_e^\dagger|s'\rangle \tilde C_{xe}^{\mu\dagger} \langle\lambda|j_x|\lambda'\rangle
\langle\lambda'|j_x^\dagger |\lambda\rangle \tilde C_{xe}^\nu  \langle s'|j_e|s\rangle  
\equiv
 {\rm Tr}\left[\tilde C_{xe}^{\mu\,\dagger} C_{xx} \tilde C_{Xe}^\nu C_{ee}  \right].
\label{eq:MmuMnu}
\end{align}
Here $C_{ee}$ is the square of the electron current part while $C_{xx}$ the square of the DM current part, 
which are defined as 
\begin{align}
C_{ee} &= {1\over s_e} 
\sum_{s,s'}
\langle s' |
\begin{pmatrix}
\mathbb{1}_e \\ 
S_e^k 
\end{pmatrix}
| s \rangle
\langle s|
( \mathbb{1}_e~ S_e^l)
| s' \rangle
 = {1\over 2} 
\begin{pmatrix}
  \tr[\mathbb{1}_e]  &  \tr[S_e^l] \\
  \tr[S_e^k] &  \tr[S_e^k S_e^l]
\end{pmatrix}
= \begin{pmatrix}
 1 & 0 \\
 0 & {1\over 4} \delta^{kl} 
\end{pmatrix},
\\%
C_{xx} &= {1\over s_x} 
\sum_{\lambda,\lambda'}
 \langle  \lambda |
\begin{pmatrix}
\mathbb{1}_x \\
S_x^m \\
\tilde{\calS}_x^{mn}   
\end{pmatrix}
| \lambda' \rangle 
\langle \lambda'| 
(\mathbb{1}_x ~ S_x^i~ \tilde{\calS}_x^{ij})
| \lambda \rangle 
 = {1\over s_x} 
\begin{pmatrix}
  \tr[\mathbb{1}_x]  &  \tr[S_x^i] & \tr[\tilde{\calS}_x^{ij}] \\
  \tr[S_x^m]  &  \tr[S_x^m S_x^i] & \tr[S_x^m \tilde{\calS}_x^{ij}] \\
  \tr[\tilde{\calS}_x^{mn}] &  \tr[\tilde{\calS}_x^{mn}S_x^i] & \tr[\tilde{\calS}_x^{mn} \tilde{\calS}_x^{ij}]
\end{pmatrix},
\end{align}
where the trace is over the spin space.
For the three kinds of DM scenarios, using the normalization of spin operators in \cref{sec:NRop}, $C_{xx}$ become, 
\begin{align}
C_{\phi\phi} = 1, \quad 
C_{\chi\chi} = 
\begin{pmatrix}
  1  &  0 \\
  0  &  {1\over 4}\delta^{m i}
\end{pmatrix}, \quad 
C_{XX}  = 
\begin{pmatrix}
  1  &  0 & 0 \\
  0  &  {2\over 3}\delta^{mi} & 0 \\
  0  &  0 & {1\over 6}(\delta^{im} \delta^{jn} + \delta^{in}\delta^{jm})
  -{1\over 9}\delta^{ij}\delta^{mn}
\end{pmatrix}.
\end{align}
In the actual calculation for the fermion and scalar cases, one can directly obtain the results from the vector case by sending the factor $2/3$ 
to  $0$ ($1/4$) for the second diagonal element in $C_{XX}$ and picking up the corresponding non-vanishing WCs according to \cref{tab:NRop} for the scalar (fermion) DM case. Therefore, the square of the matrix element becomes
\begin{align}
 \overline{|\calM_{1\to2}|^2} 
& = f_{\tt S}^* f_{\tt S} R^{00}
+ \sum_{a=1,2,3} \left(f_{\tt S}^* f_{\tt V}^a R^{a0*} + f_{\tt S} f_{\tt V}^{a*} R^{a0} \right)
+ \sum_{a,b=1,2,3}  f_{\tt V}^{a*}f_{\tt V}^{b} R^{ab}.
\end{align}
Since the dependence on $\bm{v}_0^\perp$ has been factorized out, the only three-vector available is the momentum transfer $\bm{q}$, which implies the vector and tensor-like components have the following expansion,  
\begin{align}
R^{00} \equiv a_0, \quad
R^{ab}  \equiv 
 a_1 \delta^{ab}
+ { a_2\over x_e} { q^a\over m_e} {q^b \over m_e}
- i a_3 \epsilon^{abc} {q^c \over m_e},\quad
R^{a0}  \equiv 
- i a_4 {q^a \over m_e },  
\end{align}
with the coefficients $a_{0,1,2,3,4}$ being termed the DM response functions that are listed in \cref{tab:dmRF} for the three DM scenarios. From the above we reach the general expression given in \cref{eq:M1to2sq}, 
\begin{align}
 \overline{|\calM_{1\to2}|^2} 
& = a_0 |f_{\tt S}|^2
+ a_1 |\bm{f}_{\tt V}|^2
+ {a_2 \over x_e}  \left|{\bm{q}\over m_e} \cdot \bm{f}_{\tt V} \right|^2
+ i a_3 {\bm{q} \over m_e}\cdot(\bm{f}_{\tt V}\times \bm{f}_{\tt V}^*)
+ 2\, \Im\left[a_4 f_{\tt S} \bm{f}_{\tt V}^* \cdot {\bm{q}\over m_e} \right].
\end{align}

\section{Calculation of atomic form factors}
\label{app:electronRF}

The form factors defined in \cref{eq:formfactor} can be conveniently calculated in the configuration space.
By Fourier transforming the wave-functions between the momentum and coordinate spaces, 
$\tilde \psi(\bm{k}) = \int \td^3 \bm{r}\, e^{-i \bm{k}\cdot\bm{r}} \psi(\bm{r})$, the two form factors become,
\begin{subequations}
\begin{align}
f_{1\to 2}(\bm{q})  &= 
\int \td^3 \bm{r}\, \psi^*_{k^\prime \ell^\prime m^\prime} (\bm{r}) e^{i\bm{q}\cdot\bm{r}} \psi_{n\ell m}(\bm{r}),
\\
\bm{f}_{1\to 2}(\bm{q}) &= 
\int \td^3 \bm{r}\, \psi^*_{k^\prime \ell^\prime m^\prime} (\bm{r})e^{i\bm{q}\cdot\bm{r}}
{- i\bm{\nabla} \over m_e} \psi_{n\ell m}(\bm{r}).
\label{eq:vectorff}
\end{align}    
\end{subequations}
Note the minus sign accompanying the gradient, which was missed in \cite{Catena:2019gfa}.\footnote{We got to know via electronic communications from one of the authors in \cite{Catena:2019gfa} that they included the right sign in Eq.\,(112) of their recent preprint on crystal detection of dark matter \cite{Catena:2024rym}, but no mention was made about the sign problem in \cite{Catena:2019gfa} on liquid detection. The atomic response functions and thus physics are different between the two types of detection indeed.} 
This will affect the second atomic response function $W_2$ in their paper. 

The wave functions of the initial and final state electrons are expressed as the product of a radial function and a spherical harmonic function, 
\begin{subequations}
\begin{align}
\psi_{n\ell m}(\bm{r}) &= 
R_{n\ell}(r) Y_\ell^m(\hat{\bm{r}}), \,r \equiv |\bm{r}|,\,
\hat{\bm{r}}={\bm{r} \over r},
\\
\psi_{k'\ell' m'}(\bm{r}) &= 
R_{k'\ell'}(r) Y_\ell^{m'}(\hat{\bm{r}}).
\end{align}
\end{subequations}
For the initial ground state, we take the Roothaan-Hartree-Fock (RHF) radial wave-functions which are linear combinations of the Slater-type orbitals, 
\begin{align}
R_{n\ell}(r) & = a_0^{- {3\over 2} } \sum_{j} C_{j\ell n} \left[ { (2Z_{j\ell})^{n_{j\ell}+{1\over 2} } \over [(2n_{j\ell})!]^{1\over 2} }\right]    
\left({r \over a_0}\right)^{n_{j\ell}-1} e^{- Z_{j\ell} {r\over a_0}},
\end{align}
where $a_0$ is the Bohr radius and the coefficients $(c_{j\ell n}, Z_{j\ell}, n_{j\ell})$ for the xenon and argon atoms can be found in \cite{Bunge:1993jsz}. 
For the final state in the continuum, the radial wave function can be obtained by solving the radial part of the Schr$\ddot{\rm o}$dinger equation with a modified central potential $- Z_{\rm eff}/r$ \cite{Bethe:1957ncq}, 
\begin{align}
R_{k'\ell'}(r)  & =  { 4\pi \over \sqrt{V}} (2 k' r)^{\ell'} 
{ \big|\Gamma\big(\ell'+1 - i{ Z_{\rm eff}\over k' a_0}\big) \big| 
e^{ {\pi \over2} {Z_{\rm eff} \over k' a_0} } \over (2\ell'+1)! } 
e^{-i k' r} {}_1F_1\Big(\ell'+1 + i {Z_{\rm eff}\over k' a_0}, 2\ell'+2, 2i k' r\Big).
\end{align}
There is a factor of $\sqrt{V} k'/(2\pi)^{3/2}$ difference with respect to Eq.\,(4.8) and Eq.\,(4.20) in \cite{Bethe:1957ncq} due to different normalizations given in \cite{DarkSide:2018ppu,Catena:2019gfa}, and it cancels out in the rate formula, \cref{eq:rionnl}.
The effective charge is determined by matching the energy eigenvalue to the binding energy $E_B^{n\ell}$ of the ionized orbital (i.e., the RHF eigenvalue), 
$Z_{\rm eff}^{n\ell}=n \sqrt{E_B^{n\ell}/13.6\,\rm eV}$.
While the initial-state radial function is obviously real, the final-state radial function is also real by the properties of the confluent hypergeometric function, ${}_1F_1(a,b,z)=e^z {}_1F_1(b-a,b,-z)$ and 
${}_1F_1(a,b,z)^*={}_1F_1(a^*,b^*,z^*)$.
This will be useful to prove the vanishing of the last term in \cref{eq:M1to2sq} at the end of this appendix. 

To calculate the $\bm{r}$ integrals in the form factors, it is convenient to expand also the plane wave in terms of spherical harmonics. Choosing $\bm{q}$ as the $z$ axis of $\bm{r}$, we have
\begin{align}
e^{i\bm{q}\cdot\bm{r}} 
= \sum_{L=0}^\infty \sqrt{4\pi(2L+1)} 
i^L j_L(qr)Y_L^{0}(\hat{\bm{r}}),
\quad  q \equiv |\bm{q}|,  
\label{eq:eiqrz}
\end{align}
where $j_L(qr)$ is the spherical Bessel function of the first kind with order $L$. 
The scalar form factor factorizes into a radial integral and an angular integral:
\begin{align}
f_{1 \to 2}(\bm{q})  =  \sum_{L=0}^\infty \sqrt{4\pi(2L+1)} 
i^L  I_{1,L}^{k'\ell',n\ell}(q)
\int \td\Omega_{\hat\pr}\,Y_{\ell^\prime}^{m^\prime *}(\hat{\pr})Y_{\ell}^{m}(\hat{\pr}) Y_{L}^{0}(\hat{\pr}),
\end{align}
where the radial factor is
\begin{align}
I_{1,L}^{k'\ell',n\ell}(q) &\equiv \int \td r\,r^2 R^*_{k^\prime \ell^\prime}(r) R_{n\ell}(r) j_L(qr).
\end{align}
For simplicity, we do not show explicitly the $\hat{\bm{r}}$ dependence of $Y_\ell^m$ from now on.
The angular one is evaluated by using the contraction rule for a product of two spherical harmonics \cite{Sakurai:2011zz}, 
\begin{align}
Y_{\ell_{1}}^{m_{1}} Y_{\ell_{2}}^{m_{2}}
=
\sum_{\ell=\left|\ell_{1}-\ell_{2}\right|}^{\ell_{1}+\ell_{2}} 
\sum_{m=-\ell}^{\ell} 
\sqrt{{\left(2 \ell_{1}+1\right)\left(2 \ell_{2}+1\right)\over 4 \pi(2 \ell+1)}}
\left\langle\ell_{1},0; \ell_{2},0 | \ell, 0\right\rangle\left\langle\ell_{1}, m_{1};\ell_{2}, m_{2} |\ell, m\right\rangle Y_{\ell}^{m}, 
\label{eq:YYreduction}
\end{align} 
and the orthonormal condition, so that 
\begin{align}
\int \td\Omega_{\hat\pr} 
\,Y_{\ell^\prime}^{m^\prime *}
Y_{\ell}^{m} 
Y_{L}^{0}
= (-1)^{m'}\sqrt{{\left(2 \ell'+1\right)\left(2 \ell+1\right) \over 4 \pi(2 L+1)}}
\left\langle\ell',0; \ell,0 |L, 0\right\rangle
\left\langle\ell', - m'; \ell ,m | L, 0\right\rangle. 
\label{eq:inttrihar}
\end{align}
The scalar form factor therefore becomes, 
\begin{align}
f_{1 \to 2}(\bm{q})  = (-1)^{m'}\sqrt{(2\ell+1)(2\ell'+1)}   \sum_{L=|\ell - \ell'|}^{\ell+\ell'} i^L  I_{1,L}^{k'\ell',n\ell}
\left\langle\ell',0; \ell,0 |L, 0\right\rangle
\left\langle\ell', - m'; \ell ,m | L, 0\right\rangle. 
\label{eq:scalarf12}
\end{align}
Using the orthonormal condition of the Clebsch-Gordan (CG) coefficients, its absolute square summed over $m,m'$ becomes
\begin{align}
\sum_{m,m'}| f_{1 \to 2}(\bm{q})|^2 &= 
(2\ell+1) (2\ell'+1) 
\sum_{L=|\ell - \ell'|}^{\ell+\ell'}  |I_{1,L}^{k'\ell',n\ell}(q)|^2
\langle\ell' 0; \ell, 0|L,0\rangle^2.
\end{align}
Notice that it only depends on the norm $q$ but not direction of $\bm{q}$. 

Next we consider the vector form factor in \cref{eq:vectorff}. 
The gradient of the wave function $\psi_{n\ell m}(\bm{r})$ in spherical coordinates is
\begin{align}
\bm{\nabla}\psi_{n\ell m}(\bm{r}) = 
 {\td R_{n\ell}(r)\over \td r} Y_\ell^m\,\hat{\pr}
 + {R_{n\ell} (r)\over r} \left( {\partial Y_\ell^m \over \partial\theta}\,\hat{\bm{\theta}}
 +{1\over \sin\theta} {\partial Y_\ell^m\over \partial\phi}\,\hat{\bm{\phi}}\right).
\end{align}
The spherical unit vectors $(\hat{\bm{r}}, \hat{\bm{\theta}}, \hat{\bm{\phi}})$ 
are related to the $L=1$ spherical harmonics via,  
\begin{subequations}
\begin{align}
\hat{\bm{r}} & = \sum_{w=\pm1,0} \sqrt{4\pi \over 3} Y_1^{w*}\,\hat{\bm{e}}_w^s, 
\\
\hat{\bm{\theta}} & = \sum_{w=\pm1,0} \sqrt{4\pi \over 3}{\partial Y_1^{w*}\over \partial \theta}\, \hat{\bm{e}}_w^s,
\\
\hat{\bm{\phi}} & = - \sum_{w=\pm1,0} \sqrt{{4\pi \over 3}} i\, w Y_1^{w*}(\pi/2,\phi) \hat{\bm{e}}_w^s, 
\end{align}
\end{subequations}
where $\hat{\bm{e}}_\pm^s = \mp (\hat{\bm{e}}_x \pm i \hat{\bm{e}}_y )/\sqrt{2}$, 
$\hat{\bm{e}}_0^s \equiv \hat{\bm{e}}_z$.
Together with ${\partial Y_\ell^m /\partial \phi} = i\, m\, Y_\ell^m$ and $Y_1^{w*}(\pi/2,\phi)= Y_1^{w*}/\sin\theta$, these relations imply,
\begin{subequations}
\begin{align}
\bm{\nabla}\psi_{n\ell m}(\bm{r}) & = \sqrt{4\pi \over 3}  \sum_{w=\pm 1,0}  \hat{\bm{e}}_w^s
\left(
{\td R_{n\ell} \over \td r} Y_\ell^m  Y_1^{w*}
+{R_{n\ell} \over r }X_{\ell,1}^{m,w} \right),
\\
X_{\ell,1}^{m,w} & \equiv 
{\partial Y_\ell^m \over \partial\theta}
{\partial Y_1^{w*} \over \partial \theta} 
+{ m\,w \over \sin^2\theta} Y_\ell^m  Y_1^{w*}.
\end{align}
\end{subequations}
Denoting \cite{Catena:2019gfa}
\begin{subequations}
\begin{align}
I_{2,L}^{k'\ell',n\ell}(q) & \equiv 
\int \td r\,r^2 R^*_{k^\prime \ell^\prime}(r) {\td R_{n\ell}(r)\over \td r} j_L(qr),
\\
I_{3,L}^{k'\ell',n\ell}(q) & \equiv 
\int \td r\,r R^*_{k^\prime \ell^\prime}(r) R_{n\ell}(r) j_L(qr),   
\end{align}
\end{subequations}
the vector form factor $\bm{f}_{1\to2}(\bm{q})$ is decomposed in  $\hat{\bm{e}}_{w}^{s}$, 
\begin{align}
\bm{f}_{1\to 2}\equiv \sum_{w=\pm1,0}\hat{\bm{e}}_{w}^{s} f_w^s,
\end{align} 
where,
\begin{align}
f_{w}^s  =& 
{-i\over m_e}  \sum_{L=0}^\infty i^L\sqrt{(4\pi)^2(2L+1)\over 3}
\nonumber
\\
&\times\left( I_{2,L}^{k'\ell',n\ell} 
\int \td\Omega_{\hat\pr} ~Y_{\ell^\prime}^{m^\prime *} Y_\ell^m Y_1^{w*}  Y_{L}^{0} 
+I_{3,L}^{k'\ell',n\ell} 
\int \textbf{}d\Omega_{\hat\pr} ~Y_{\ell^\prime}^{m^\prime *} X_{\ell,1}^{m,w} Y_{L}^{0}\right).
\end{align}

To proceed further we need the derivative relations \cite{Khersonskii:1988krb},
\begin{subequations}
\begin{align}
{\partial Y_\ell^m \over \partial\theta} 
&=m {\cos\theta\over \sin\theta} Y_\ell^m 
+ \sqrt{(\ell - m)(\ell+m+1)}Y_\ell^{m+1} e^{-i\phi},
\\
{\partial Y_\ell^m \over \partial\theta} 
& = -m {\cos\theta\over \sin\theta} Y_\ell^m 
- \sqrt{(\ell + m)(\ell-m+1)}Y_\ell^{m-1} e^{i\phi}, 
\end{align}
\end{subequations} 
so that we obtain 
\begin{subequations}
\begin{align} 
X_{\ell,1}^{m,w=\pm1}& = w\, m Y_\ell^m  Y_1^{w*}  
+ \sqrt{(\ell + w\, m)(\ell- w\, m +1)/2}\,  Y_\ell^{m- w } Y_1^{0*},
\\   
X_{\ell,1}^{m,w=0}& = \sum_{w=\pm1} \sqrt{(\ell + w\, m)(\ell- w\, m +1)/2}\,  Y_\ell^{m- w } Y_1^{-w*}.
\end{align}
\end{subequations}
Using the above results, the contraction rule in \cref{eq:YYreduction}, and the orthonormal condition in \cref{eq:inttrihar}, we obtain the angular integrals in $f_w^s$:
\begin{subequations}
\begin{align}
\int d\Omega_{\hat\pr} Y_{\ell^\prime}^{m^\prime *} Y_\ell^m   Y_1^{w*} Y_{L}^{0} 
& = (-1)^{m'} \sqrt{ { 3 \left(2 \ell'+1\right) \left(2 \ell+1\right)\over (4 \pi)^2 (2 L+1)}} 
P_{\ell',m';\ell,m}^{L,w}, 
\\
\int d\Omega_{\hat\pr} Y_{\ell^\prime}^{m^\prime *} Y_{L}^{0} X_{\ell, 1}^{m,w} 
& = (-1)^{m'} \sqrt{ { 3 \left(2 \ell'+1\right) \left(2 \ell+1\right)\over (4 \pi)^2 (2 L+1)}} 
Q_{\ell',m';\ell,m}^{L,w},
\end{align}    
\end{subequations}
with 
\begin{align}
P_{\ell',m';\ell,m}^{L,w}\equiv& (-1)^{w}
\sum_{j}
\left\langle 1,0; j,0 | L, 0\right\rangle
\left\langle\ell',0; \ell,0 |j, 0\right\rangle
\left\langle 1, -w; j , w | L, 0 \right\rangle 
\left\langle\ell',- m'; \ell, m |j,w \right\rangle,
\\
Q_{\ell',m';\ell,m}^{L,w=\pm1}\equiv& -
\sum_{j}
\left\langle 1,0; j,0 | L, 0\right\rangle
\left\langle\ell',0; \ell,0 |j, 0\right\rangle 
\Big[
w\, m\, \left\langle 1, -w; j , w | L, 0 \right\rangle 
\left\langle\ell',- m'; \ell, m |j,w \right\rangle
\nonumber
\\
& 
-  \sqrt{(\ell + w\, m)(\ell- w\, m +1)/2} 
\left\langle 1, 0; j , 0 | L, 0 \right\rangle 
\left\langle\ell',- m'; \ell, m -w |j,0 \right\rangle
\Big],
\\
Q_{\ell',m';\ell,m}^{L,w=0}\equiv& - \sum_{j}
\left\langle 1,0; j,0 | L, 0\right\rangle
\left\langle\ell',0; \ell,0 |j, 0\right\rangle
 \left\langle 1, -1; j , 1 | L, 0 \right\rangle
 \left[
 \sqrt{(\ell +  m)(\ell-  m +1)/2}  
 \right.
\nonumber
\\
&\left.
\times \left\langle\ell', m'; \ell, 1 - m | j, 1 \right\rangle
 +  \sqrt{(\ell -  m)(\ell + m +1)/2}  \left\langle\ell', - m'; \ell, 1 +m | j, 1 \right\rangle
\right].
\end{align}
We finally obtain 
\begin{align}
f_{w}^s 
= { (-1)^{m'} \over m_e}  \sqrt{\left(2 \ell+1\right)\left(2 \ell'+1\right) }
\sum_{L=||\ell -\ell'|- 1|}^{\ell'+\ell+1}i^{L-1}
\left(
I_{2,L}^{k'\ell',n\ell} P_{\ell',m';\ell,m}^{L,w}
+I_{3,L}^{k'\ell',n\ell} Q_{\ell',m';\ell,m}^{L,w}  \right) .
\label{eq:fspw}
\end{align}
Note that the real coefficients $P,\,Q$ satisfy the properties, 
$P_{\ell',m';\ell,m}^{L,w}=P_{\ell',-m';\ell,-m}^{L,-w}$ and  $Q_{\ell',m';\ell,m}^{L,w}=Q_{\ell',-m';\ell,-m}^{L,-w}$, 
leading to $f_{-w}^s = f_w^s|_{(m',m)\to(-m',-m)}$.

Now we can understand why the terms with $a_{3,4}$ vanish once $m,m'$ are summed over. For $f_w^s$ the non-vanishing contribution requires $w= m-m'=\pm 1,0$ by the properties of CG coefficients. 
However, the second CG coefficient of $f_{1\to2 }$ in \cref{eq:scalarf12} implies $m=m'$, 
which leads to 
\begin{align}
\sum_{m,m'}  f_{1\to2 } \bm{f}_{1\to 2}^* =
\hat{\bm{e}}_0^{s*} \sum_{m,m'}  f_{1\to 2} f_{0}^{s*}
 = \hat{\bm{q}}\sum_{m,m'}  f_{1\to 2} f_{0}^{s*}. 
\end{align}
Thus, the first term on the right-hand side of \cref{eq:a3coeff} vanishes due to $\hat{\bm{q}}\cdot (\bm{q}\times \bm{v}_0^\perp)=0$. 
Since 
$i \hat{\bm{q}}\cdot (\bm{f}_{1\to 2} \times \bm{f}_{1\to 2}^*) =\sum_{w=\pm1} w f_{w}^s f_{w}^{s*} =|f_{+1}^{s}|^2 -|f_{-1}^{s}|^2$,
by \cref{eq:fspw} and its property under $w\to -w$, it is easy to see $\sum_{m,m'}|f_{-1}^{s}|^2 =\sum_{m,m'} |f_{+1}^{s}|^2 $ and thus the vanishing of the second term in \cref{eq:a3coeff}.
This finishes the proof that the term with $a_3$ has no contribution. 
The vanishing of the term with $a_4$ is 
based on the assumption of real $a_4$ or real WCs.
The first term on the right-hand side of \cref{eq:a4coeff} is obvious, while the second term 
comes from the fact that the initial and final radial wave-functions are real and the power of $i$ in nonvanishing $f_{1\to 2} \bm{f}_{1\to 2}^* $ is always an even integer number,
\footnote{
For the CG coefficient  $\langle j_1,0; j_2,0| j,0\rangle$, a non-vanishing result requires $j+j_1+j_2$ to be an even integer. Thus, from \cref{eq:scalarf12}, the power ($L$) of $i$ in $f_{1\to 2} $  can only take the values $\{|\ell- \ell'|,|\ell- \ell'|+2,|\ell- \ell'|+4,  \cdots, \ell+\ell'\}$ to make it nonzero. Similarly, from \cref{eq:fspw}, the power ($L-1$) of $i$ in $f_w^s$ takes the values $\{||\ell- \ell'|-1|-1,||\ell- \ell'|-1|+1,\cdots, \ell+\ell'\}$. Both powers of $i$ in $f_{1\to2}$ and $f_w^s$ have the same even/odd property, resulting in an even power of $i$ in $f_{1\to 2}f_0^{s*} $. }
leading to $\sum _{m,m'} \Im \left[ f_{1\to 2} \bm{f}_{1\to 2}^* \cdot \hat{\bm{q}} \right] =
\sum _{m,m'} \Im \left[ f_{1\to 2} f_0^{s*} \right]=0$. 

By the orthonormality of the spherical unit vectors and the above equation, 
$ \bm{q}\cdot\bm{f}_{1\to 2} = q f_{0}^s$ and 
$ \sum _{m,m'} f_{1\to 2} \bm{f}_{1\to 2}^* \cdot\bm{q}/m_e = \sqrt{x_e} \sum _{m,m'} f_{1\to 2} f_{0}^{s*}$,
the last three response functions in \cref{eq:W1to4} become, 
\begin{subequations}
\begin{align}
W_1 & = N_{k'}  \,\sum_{m ,\ell'} \left|f_{1\to 2}\right|^2_{m'=m}, 
\\
W_2 & = N_{k'}  \,\sqrt{x_e} \sum_{m ,\ell'}  f_{1\to 2} f_{0}^{s*} \big|_{m'=m},
\\
W_3 &  =N_{k'}  \,\sum_{m ,\ell'} \left(\left|f_0^s\right|^2_{m'=m} +  2 \left|f_{+1}^s\right|^2_{m'=m-1}\right) , 
\\
W_4 &= N_{k'}  \, x_e \sum_{m ,\ell'} \left|f_0^s \right|^2_{m'=m} ,
\end{align}
\end{subequations}
where $N_{k'} \equiv V{4 k'^3/ {(2 \pi)^3}}$. Then, $\widetilde W_2$ defined in \cref{eq:W012tilde} can be exactly reorganized into a simple form,  
\begin{align}
 \widetilde W_2  = N_{k'} \sum_{m ,\ell'}
\left| { y_e \over \sqrt{x_e}}f_{1\to 2}- f_0^s\right|^2_{m'=m}.
\label{eq:W2t}
\end{align}
Neglecting the tiny correction due to the DM mass and the square term of $\Delta E_{1\to2}/m_e$ as in \cref{eq:Wtildeapp}, 
$\widetilde W_1$ becomes, 
\begin{align}
\widetilde W_1 \approx \widetilde W_2 + 2 N_{k'} \sum_{m ,\ell'}\left|f_{+1}^s\right|^2_{m'=m-1}.
\label{eq:W1t}
\end{align}

\begin{figure}
\includegraphics[width=0.99 \textwidth]{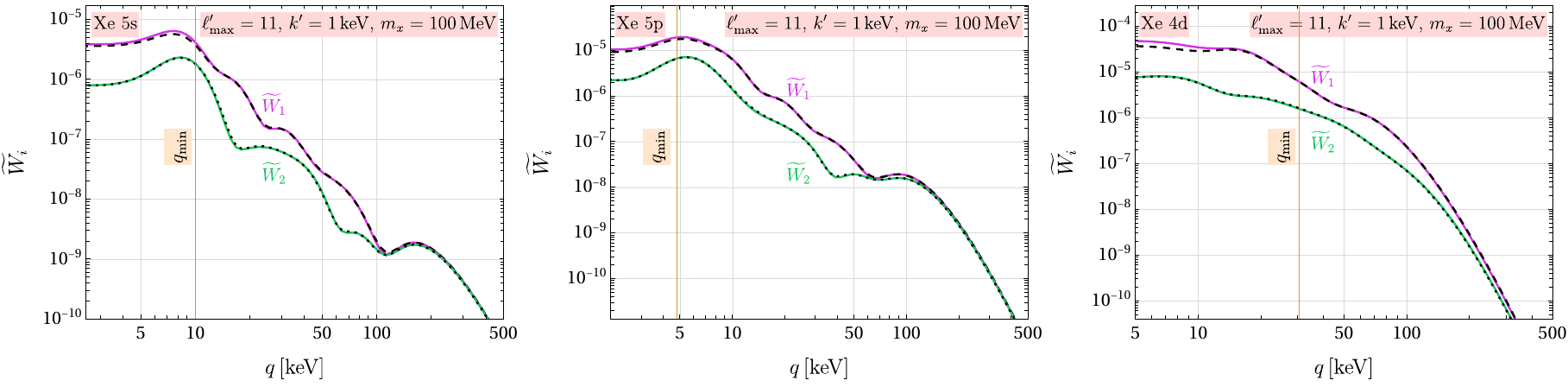}
\caption{Comparison of $\widetilde W_{1,2}$ between \cref{eq:W1t,eq:W2t} (solid color curves) and exact (dashed and dotted black curves) results in \cref{eq:W012tilde} for several atomic orbitals of a xenon target. Notice that the DM mass $m_x$ is only relevant to the location of $q_{\rm min}$ and has a negligible effect on $\widetilde W_{1,2}$.
}
\label{fig:W12tilde}
\end{figure} 
\begin{figure}
\includegraphics[width=0.47 \textwidth]{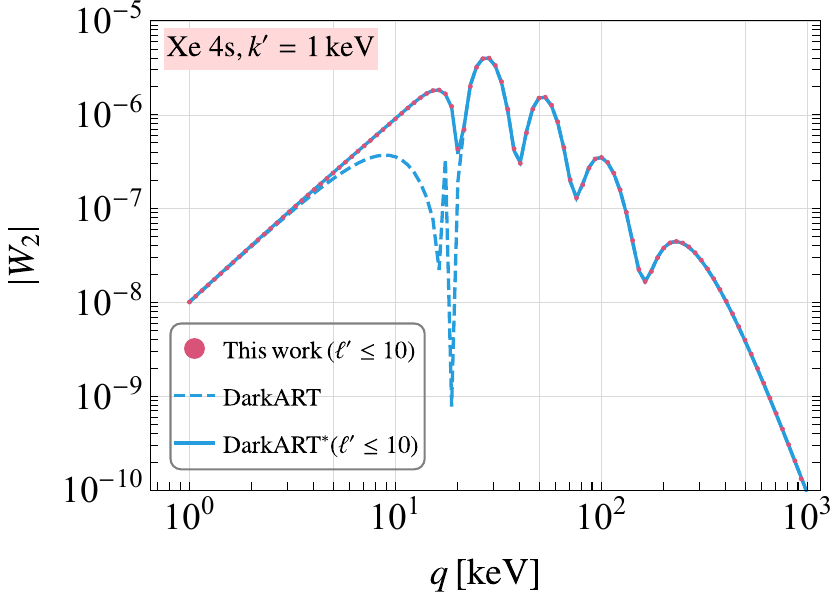}
\includegraphics[width=0.45 \textwidth]{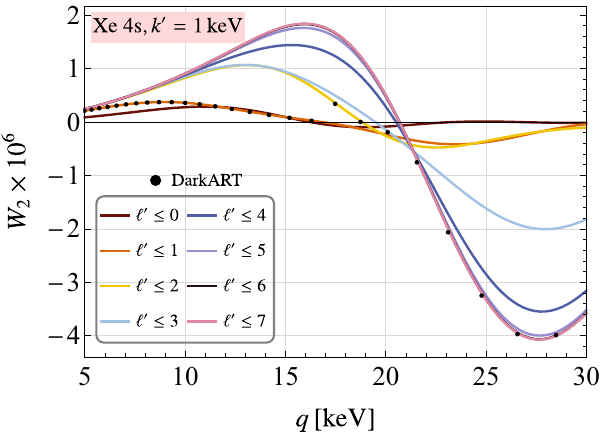}
\caption{
$W_2$ for the 4s orbit of a xenon target with $k' = 1\,\rm keV$. \textbf{Left:} The red dots (blue dashed curve, blue solid curve) show the result by our {\em Mathematica} code summed over $\ell' \leq 10$ (by the original {\tt DarkART}, by the {\tt DarkART} summed over  $\ell' \leq 10$). 
\textbf{Right:}  Various curves showing our results summed over $\ell' \leq 1,2,\cdots,7$ are compared with that (black dots) by the original {\tt DarkART}.
}
\label{fig:W2-check}
\end{figure} 

The above formulas are very convenient for numerical evaluation. In \cref{fig:W12tilde}, we show by solid curves $\widetilde W_{1,2}$ based on \cref{eq:W1t,eq:W2t} and evaluated by our {\em Mathematica} code for several atomic orbitals with a fixed momentum of ionized electron $k'=1\,\rm keV$. For comparison, we also show the exact results (dashed and dotted curves) according to our \cref{eq:W012tilde} where $W_{1-4}$ are evaluated by the code {\tt DarkART} provided in \cite{DarkART,Catena:2019gfa} but with a manually corrected minus sign for $W_2$. The vertical line labeled by $q_{\rm min}$ means that only the region on its right-hand side will contribute to the scattering rate. As can be seen, 
our calculation for $\widetilde W_2$ based on \cref{eq:W1t} is consistent with the one from \cref{eq:W012tilde}, while for $\widetilde W_1$ the approximation is rather good with only a minor difference in the lower region $q < q_{\rm min}$, which does not contribute to the scattering rate anyway. 
As a final side remark, we mention an issue concerning the truncation of $\ell'$ in the sum over the final electron orbitals in \cref{eq:W1to4}. We find that for some specific orbits such as 4s the $W_2$ calculated by {\tt DarkART} is truncated too early, leading to a distinct discrepancy with our result, as shown in \cref{fig:W2-check}. But when we postpone the truncation in {\tt DarkART} until $\ell' = 10$, we were able to achieve a high level of consistency with our result. Therefore, to be conservative, we truncate at $\ell' = 30$ when we employ the high-speed {\tt DarkART} (written in {\tt C++}) to obtain the atomic response functions as the input for our constraint calculation.

\bibliography{refs_paper.bib}{}
\bibliographystyle{JHEP}

\end{document}